\documentclass[aps,prl,twocolumn,superscriptaddress]{revtex4-2}
\usepackage{chngcntr}
\usepackage{graphicx}
\usepackage{color}
\usepackage{amsmath}
\usepackage{enumitem}
\usepackage{amssymb}
\usepackage{hyperref}
\usepackage{cancel}
\usepackage{ulem}
\usepackage{multirow}
\usepackage{CJK}
\usepackage{verbatim}
\usepackage{dcolumn}
\usepackage{bm}
\usepackage{subfigure}
\usepackage{xcolor,float}

\newcommand\startsupplement{%
       \newpage\clearpage
       \setcounter{secnumdepth}{2}
       \setcounter{table}{0}
       \renewcommand{\thetable}{S\arabic{table}}
       \setcounter{figure}{0}
       \renewcommand{\thefigure}{S\arabic{figure}}
       \setcounter{equation}{0}
       \renewcommand{\theequation}{S\arabic{equation}}
       \setcounter{section}{0}
       \renewcommand{\thesection}{Section \Roman{section}}
       \renewcommand{\thesubsection}{\Roman{section}. \Alph{subsection}}
    }

\usepackage{xcolor}
\definecolor{orange}{rgb}{1,0.5,0}
\definecolor{darkblue}{rgb}{0.,0.,0.4}
\definecolor{darkred}{rgb}{0.5,0.,0.}

\usepackage{comment}

\begin{document}

\title{ Conformal Operator Flows of the Deconfined Quantum Criticality from $\mathrm{SO}(5)$  to $\mathrm{O}(4)$ }

\author{Shuai Yang}	
\altaffiliation{The two authors contributed equally to this work.}
\affiliation{Department of Physics and State Key Laboratory of Surface Physics, Fudan University, Shanghai 200433, P.R. China}
\author{Liang-dong Hu}
\altaffiliation{The two authors contributed equally to this work.}
\author{Chao Han}
\author{W. Zhu}
\email{zhuwei@westlake.edu.cn}
\affiliation{Institute of Natural Sciences, Westlake Institute for Advanced Study, Hangzhou 310024, China}
\affiliation{Department of Physics, School of Science, Westlake University, Hangzhou 310030, China }

\author{Yan Chen}
\email{yanchen99@fudan.edu.cn}
\affiliation{Department of Physics and State Key Laboratory of Surface Physics, Fudan University, Shanghai 200433, P.R. China}
\affiliation{Shanghai Branch, Hefei National Laboratory, Shanghai 201315, P.R. China}

\date{\today}

\begin{abstract}

The deconfined quantum critical point (DQCP), which separates two distinct symmetry-broken phases, was conjectured to be an example of (2+1)D criticality beyond the standard Landau-Ginzburg-Wilson paradigm. However, this hypothesis has been met with challenges and remains elusive. Here, we perform a systematic study of a microscopic model realizing the DQCP with a global symmetry tunable from $\mathrm{SO}(5)$ to $\mathrm{O}(4)$. Through the lens of fuzzy sphere regularization, we uncover the key information on the renormalization group flow of conformal operators. 
We reveal O(4) primaries decomposed from original SO(5) primaries by 
tracing conformal operator content and identifying the ``avoided level crossing'' in the operator flows. 
In particular, we find that
the existence of a scalar operator, in support of the nature of pseudo-criticality, remains relevant, persisting from $\mathrm{SO}(5)$ to $\mathrm{O}(4)$ DQCP.
This work not only uncovers the nature of O(4) DQCP but also demonstrates that the fuzzy sphere scheme offers a unique perspective on the renormalization group flow of operators in the study of critical phenomena.

\end{abstract}
	
\maketitle

\textit{Introduction.---} 
The deconfined quantum critical point (DQCP) was initially proposed to describe a direct phase transition between the N\'eel and valence-bond-solid (VBS) phase \cite{Read-and-Sachdev-PhysRevB.42.4568,Senthil-doi:10.1126/science.1091806,Senthil-PhysRevB.70.144407}. It attracts widespread attention because the DQCP serves as a playground for exotic concepts like emergent gauge fields and deconfined elementary excitations \cite{Senthil-doi:10.1126/science.1091806, Senthil_review}, which is absent within the conventional Landau-Ginzberg-Wilson paradigm. Recent theoretical interests have concentrated on the exploration of the DQCP in various systems, such as the J-Q model \cite{Sankvik2007, Sandvik2009}, the 3D loop model \cite{NahumSO5, NahumO4}, and the Heisenberg model on various lattices \cite {JongYeonLeePhysRevX.9.041037, WenYuanLiuPhysRevLett.133.026502,LingWangPhysRevB.105.L060409,XueFengZhang2018,Kaul2016,YQQin2017} and the fermionic models \cite{YLiu2019,Zixiangli2019}, as well as the ultimate fate of the DQCP flowing to a conformally invariant fixed point \cite{Senthil_review,ChongWang-PhysRevX.7.031051,NahumO4}.
However, despite intense studies, the precise nature of the DQCP remains controversial.
For instance, there is accumulated evidence that the DQCP is not a continuous transition: 1) The discrepancy of the finite-size scaling cannot reconcile with regular ansatz of second-order transitions \cite{FJJiang2008, Troyer2008, KChen2013, HShao2016}; 2)
The extracted critical exponents are incompatible with bounds from conformal bootstrap calculations \cite{Nakayama-PhysRevLett.117.131601, Li_2022O4DQCP}; 3) The calculations of order parameters experience a discontinuity at the transition point \cite{NahumO4, Emidio_2023, BZhao2019}; 4) Quantum entanglement measurements identify indirect evidence for the symmetry breaking effect \cite{ZDeng2024}. 
These findings cast doubt on the nature of DQCP and call for a dissection of the operator content of DQCP; however, this is out of reach using traditional methods.  

Very recently, a novel framework for studying 3D conformal field theory (CFT)---fuzzy sphere regularization---has been introduced \cite{ZHHHH2022}. It can directly expose the underlying CFT algebra and conformal operator content by utilizing the celebrated state-operator correspondence \cite{Cardy1984, Cardy1985a}. So far, it has been successfully applied to Landau symmetry broken transitions  \cite{ZHHHH2022,han2023conformaloperatorcontentwilsonfisher,lauchli2025exactdiagonalizationmatrixproduct, SYang2025,voinea2024}, Lee-Yang transition \cite{miro2025flowingisingmodelfuzzy,fan2025simulatingnonunitaryyangleeconformal,cruz2025yangleequantumcriticalityvarious}, defect and surface criticalities \cite{hu2023defect, Zhou2025surface}, transitions involving topological orders \cite{voinea2024}. It is also powerful on exploring microscopic conformal generators \cite{fan2024,fardelli2024}, conformal correlators \cite{Four_Han2023}, renormalization group (RG) monotonic quantities \cite{hu2024F,Zhou2024gfunction}. 
Especially when applying it to the $\mathrm{SO}(5)$ DQCP problem \cite{zhou2024mathrmso5deconfinedphasetransition,zhou2024newseries3dcfts}, the fuzzy sphere regularization facilitates the discernment of underlying CFT algebra, including the emergent conformal symmetry and identification of a list
of conformal primary fields (e.g. the $\mathrm{SO}(5)$ order parameter, the rank-2 tensor operator, monopole operators) \cite{zhou2024mathrmso5deconfinedphasetransition}. Crucially, it discovers a relevant singlet primary field $S$, which was previously unknown. The appearance of this salient, relevant primary inevitably leads to instability towards a weakly first-order transition \cite{zhou2024digitalso5deconfinedphasetransition, Gorbenko2018a, Gorbenko2018b, Kaplan2009}. 
The conformal bootstrap calculation has independently verified the existence of such a singlet operator $S$ \cite{ShaiChester-PhysRevLett.132.111601} and Monte Carlo calculations on the lattice J-Q model \cite{takahashi2024}.  These progresses on the $\mathrm{SO}(5)$ DQCP motivate us to revisit the problem of the
$\mathrm{O}(4)$ DQCP \cite{Senthil-PhysRevB.70.144407}, a different class of putative deconfined phase transitions involving four-component soft order parameter:  Is a hypothetical O(4)-symmetric fixed point plausible, which was previously endorsed by the duality \cite{ChongWang-PhysRevX.7.031051,jian2017emergentsymmetrytricriticalpoints, Lu2021}? Does the O(4) DQCP share the same physics with $\mathrm{SO}(5)$ case?

In this work, we address the aforementioned questions by studying a microscopic model with the Neel-to-VBS transition whose global symmetry can be continuously tuned from $\mathrm{SO}(5)$ to $\mathrm{O}(4)$. With the help of fuzzy sphere regularization, we extract crucial conformal data from the operator spectra and trace the running flows of operator content. The main findings include: 1) 
By reducing symmetry, the $\mathrm{SO}(5)$ primaries split and decompose into $\mathrm{O}(4)$ fields.
2) Two operators with the same symmetry experience a ``avoided level crossing'' in the RG flow by exchanging their operator content.  
3) The scalar primary $S$ remains relevant in the RG flow, which implies that the $\mathrm{O}(4)$ DQCP is not a genuine fixed point, and may share the same pseudo-critical behavior as the $\mathrm{SO}(5)$ case \cite{zhou2024mathrmso5deconfinedphasetransition}. 
These findings are crucial not only for the $\mathrm{O}(4)$ DQCP but also serve as the first example of RG flows of conformal operators in the microscopic study of critical phenomena.

\begin{figure}[t]
\includegraphics[width=0.99\linewidth]{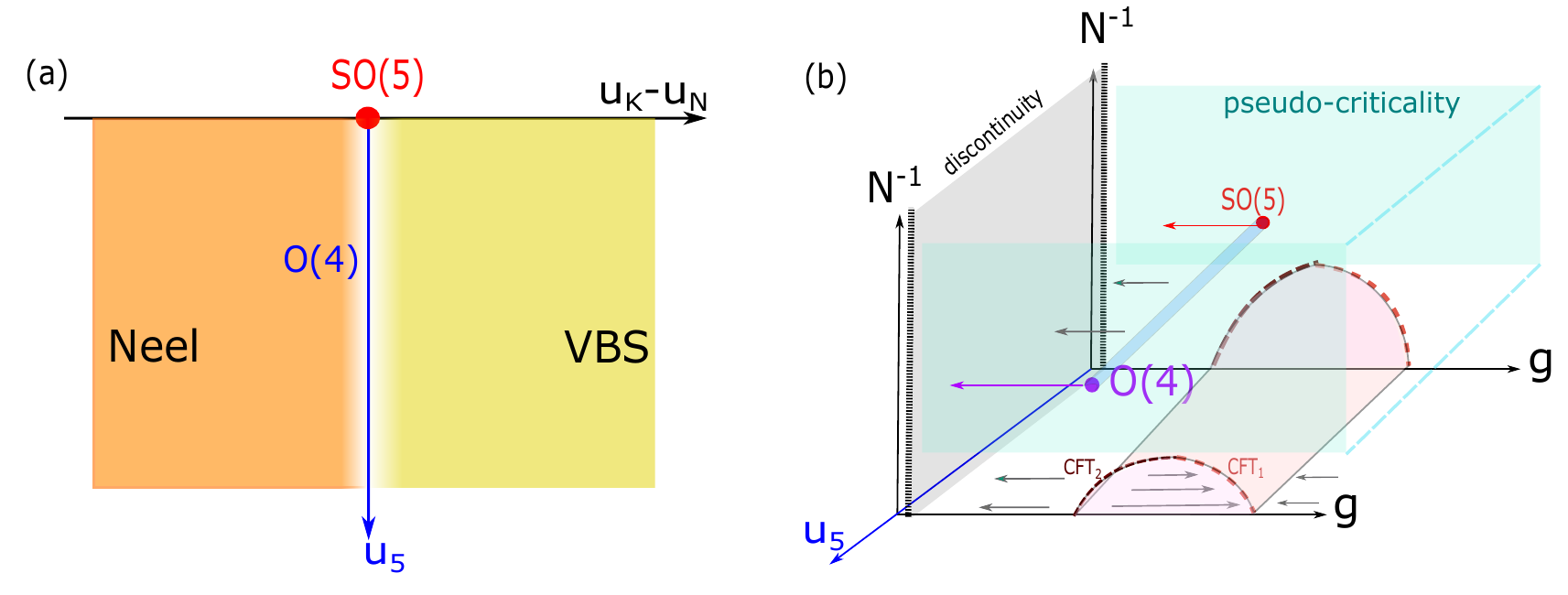}
	\caption{ 
    (a) Schematic phase diagram of the model in Eq. \ref{eq:ham_real}, where the N\'eel-to-VBS transition is tuned by an additional parameter $u_5$  to control the global symmetry at the transition, i.e. the $\mathrm{SO}(5)$ is realized at $u_5=1$ while it reduces to $\mathrm{O}(4)$ for $u_5<1$. (b) Putative RG diagram  of a 3D NL$\sigma$M with a WZW term extending by $u_5$ ($N$ is number of fermion flavor and $g$ is NL$\sigma$M coupling strength). The black-shaded plane demonstrates the discontinuity fixed point, representing a symmetry-broken fixed point. In the pseudo-critical region (green area), the RG flow runs slowly and eventually reaches the discontinuity fixed point.   
    } 
	\label{fig:drawing}
\end{figure}

\textit{Model and method.---}
We consider a 4-flavor interacting fermion model \cite{Zaletel-PhysRevB.98.235108,zhenjiuwang-PhysRevLett.126.045701,zhou2024mathrmso5deconfinedphasetransition}:
\begin{equation}
\begin{split}
    H = \int d\boldsymbol{r}_1d\boldsymbol{r}_2&\delta \left( \boldsymbol{r}_{1}-\boldsymbol{r}_2 \right)\Biggl[ 
     \boldsymbol{n}_0\left( \boldsymbol{r}_1 \right) \boldsymbol{n}_0\left( \boldsymbol{r}_2 \right) \\
    & \hphantom{u_0} -\sum_{i=1}^5{u_i\boldsymbol{n}_i\left( \boldsymbol{r}_1 \right) \boldsymbol{n}_i\left( \boldsymbol{r}_2 \right)} \Biggr],
\end{split}
\label{eq:ham_real}
\end{equation}
where  density operators are $\boldsymbol{n}_i=\boldsymbol{\psi}^\dagger \Gamma^i\boldsymbol{\psi}$, $\Gamma^0=\mathbb{I}\otimes\mathbb{I}$, and $\Gamma^{i=1,\cdots,5}=\{\tau_x\otimes\mathbb{I},\tau_y\otimes\mathbb{I},\tau_z\otimes\vec{\sigma}\}$ represent the Clifford algebra of the SO(5) group.
We set coupling strength of the N\'eel (VBS) order parameter as $u_{3,4,5}=u_N (u_{1,2}=u_K)$, so that the N\'eel (VBS) phase is energetically favored by $u_{N}>u_K (u_{N}<u_K)$. Especially, the phase transition occurs at the exact $\mathrm{SO}(5)$ point $u_{N}=u_{K}$ (see Fig. \ref{fig:drawing}(a)).
To obtain the O(4) critical theory, it is necessary to introduce an explicit symmetry-breaking perturbation that reduces the symmetry from SO(5) to O(4). This perturbation is achieved by adjusting the parameter $u_5<u_{1,2,3,4}$, i.e., the fifth vector $n_5$ becomes disfavoured, while the remaining four vectors retain rotational freedom within a 4-dimensional subspace. Equivalently speaking, we preserve 6 out of the original 10 SO(5) generators $L^{ij}=-\frac{i}{2}[\Gamma^i,\Gamma^j]$ to construct an O(4)-symmetric model.

It is expected that the low-energy part of this model will flow to the non-linear $\sigma$-model (NL$\sigma$M)  with a level-1 Wess-Zumino-Witten (WZW) term \cite{XiaoHu2005,Senthil-PhysRevB.70.144407,Lee2015}.  
For the SO(5) case, the RG phase diagram strongly depends on the number of fermion flavor $N$
\cite{Ma2020Theory,Nahum2020WZW}: For $N>N_c$, two real CFT fixed points exist, while a pseudo-critical behavior \cite{Gorbenko2018a,Gorbenko2018b,Kaplan2009} occurs for $N<N_c$. In the pseudo-critical region, the RG flows converge to a discontinuity fixed point.
By introducing an anisotropy term $u_5$, it may support a similar RG phase diagram for the O(4) case (see Fig. \ref{fig:drawing}(b)). Apart from large-N cases \cite{largeN-easyplaneDQCP-PhysRevLett.118.187202}, 
non-perturbative calculations supports the O(4) easy-plane Neel-to-VBS transition falling into the pseudo-critical regime \cite{BZhao2019,NahumO4}.

\begin{figure}[b]
\includegraphics[width=0.97\linewidth]{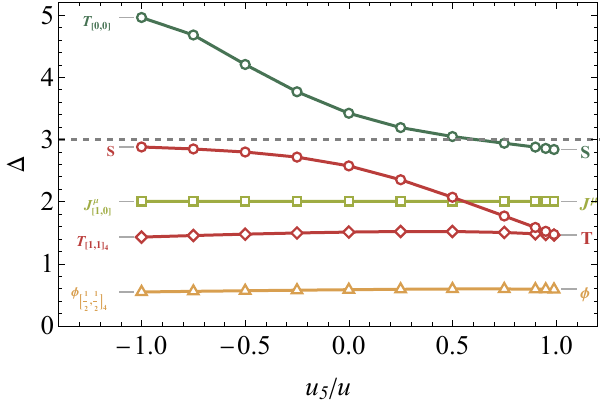}
	\caption{ 
    Evolution of low-lying operators with varying $u_5/u$. $u_5/u=1$ is the SO(5) point \cite{zhou2024mathrmso5deconfinedphasetransition}. By reducing $u_5$, the SO(5) primaries split and continuously evolve into O(4) fields. 
    Here, $[j,k]$ denotes the irreducible representations of two SU(2) subgroups within SO(4) \cite{SM}. Only parity-even operators are shown here, and the full spectral flow is presented in Fig. S2 in Supple. Mat. \cite{SM}. 
    } 
	\label{fig:flow_even}
\end{figure}

Next, we use the fuzzy sphere scheme to solve the quantum Hamiltonian Eq. \ref{eq:ham_real} \cite{ZHHHH2022} numerically. 
In specific, we project it into the spherical lowest Landau level (LLL) \cite{SM}: $\boldsymbol{\psi}(\boldsymbol{r})=\frac{1}{\sqrt{N_{\rm{o}}}}\sum_{m=-s}^{s}\Bar{Y}_{sm}^{(s)}(\boldsymbol{r})\boldsymbol{c}_m$, where ${Y}_{sm}^{(s)}(\boldsymbol{r})$ is monopole harmonics with $s$ monopole charge placed at the center of sphere and the number of degenerated Landau orbitals is $N_{\rm{o}}=2s+1$. 
Throughout the paper, we consider fermions with half-filling of the LLL. 
We perform an exact diagonalization (ED) calculation to obtain the low-lying energy spectra. According to the state-operator correspondence \cite{Cardy1984,Cardy1985a} for a CFT on the spherical geometry, each eigenstate has a one-to-one correspondence with a CFT operator and its eigenenergy $\delta E_n=E_n-E_0$ is proportional to the scaling dimension $\{\Delta_n\}$ as $\delta E_n = \frac{v}{R}(\Delta_n-\Delta_0)$, where $v$ is a model-dependent velocity and $R$ is the radius of the sphere. Each eigenstate is labeled by the total angular momentum $\ell$ of $\rm{SO(3)}$ spatial rotation symmetry and the $\rm{O(4)}$ Casimir number (see Supple. Mat. \cite{SM}). 
Since the conserved current  $J^\mu(\ell=1)$ and energy-momentum tensor ${T}^{\mu\nu}(\ell=3)$ operators exist, we use the following recipe \cite{zhou2024mathrmso5deconfinedphasetransition} to search the potential O(4) conformal fixed point: The optimal critical value is determined by the condition that scaling dimension of the energy-momentum tensor operator $\mathcal{T}^{\mu\nu}$ is set to 3 and the scaling dimension corresponding to the $J^\mu$-operator approaches 2 simultaneously.  
For each parameter $u_5/u\in [-1,1]$, we repeat the above process to determine the critical point $u_c$.

\begin{figure}[t]
\raggedleft
\includegraphics[width=0.93\linewidth]{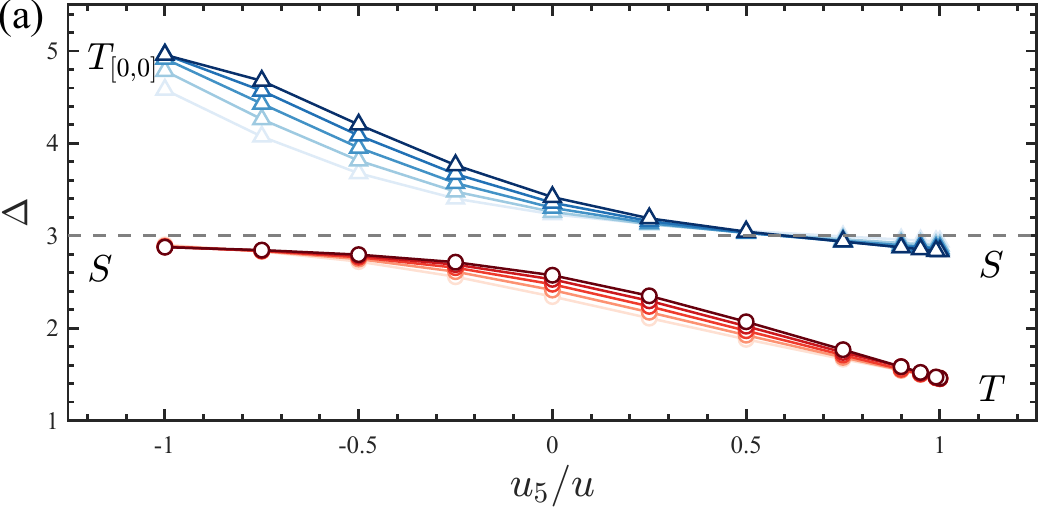}
\includegraphics[width=0.95\linewidth]{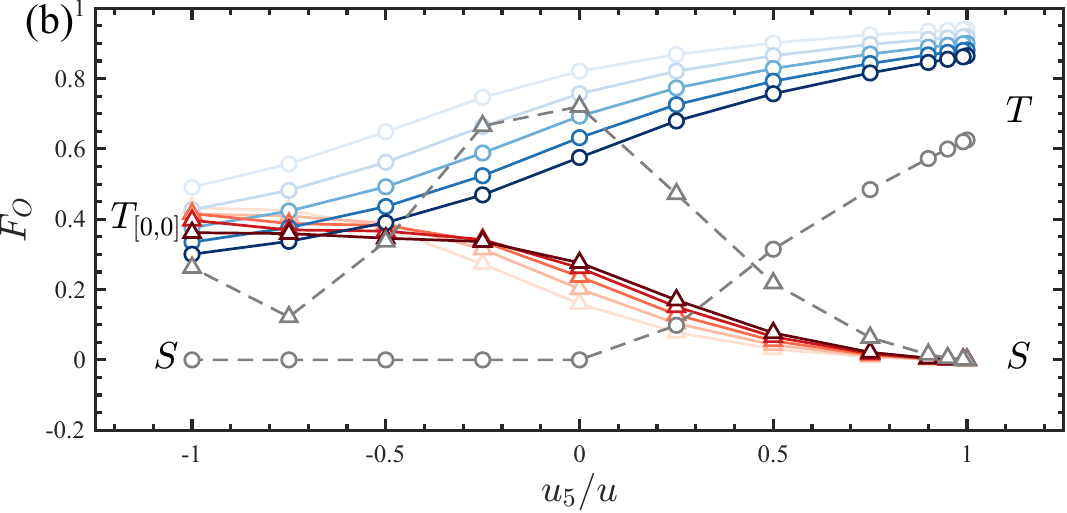}
	\caption{  Avoided level crossing between $S$ and $T_{[0,0]}$. (a) Evolution of scaling dimensions of two lowest singlets relating to $S$ and $T_{[0,0]}$. 
    (b) Operator content of $S$ and $T_{[0,0]}$ from the inner product $F_O=\langle(S, T_{[0,0]}) |O=T_{55} | I\rangle/|T_{55}|I\rangle|$, where a crossing behavior thereby suggests the operator content switching with each other. 
    Different colors correspond to system sizes with $N_o=6-10$ (colors from light to dark), and the dashed lines (gray symbols) in (b) represent the finite-size extrapolation results. 
    } 
	\label{fig:overlap}
\end{figure}

\textit{Numerical results.---}
We first analyze how the CFT fields evolve when the global symmetry is explicitly broken from SO(5) to O(4).
In Fig. \ref{fig:flow_even}, we observe that primaries with representation under the O(4) group are decomposed from original SO(5) primaries. For example,
the O(4) parity-odd vector $\phi_{[\frac{1}{2},\frac{1}{2}]}$  is decomposited from the SO(5) primary $\phi$,
and the O(4) current $J^{\mu}_{[1,0]}(J^{\mu}_{[0,1]}$) is from the SO(5) current $J^\mu$. Here, $[j,k]$ denotes the irreducible representations of the two SU(2) subgroups of SO(4) (please see Supple. Mat. \cite{SM}).
In Fig. \ref{fig:flow_even}, we only show the parity-even operators and the full spectral flow. Please see Supple. Mat. \cite{SM}. 

By inspecting these flows in detail in Fig. \ref{fig:flow_even}, we identify most of the low-lying parity-even fields evolve very smoothly from SO(5) to O(4). 
Moreover, by reducing $u_5$, two scalar operator $S$ and $T_{[0,0]}$ show sizable changes with the anisotropy term, 
i.e., the relative gap between operator $S$ and $T_{[0,0]}$ initially decreases and then increases, ultimately attaining a minimal value around $u_5\approx 0.1$ (Fig. \ref{fig:overlap}(a)). This observation strongly suggests an ``avoided crossing'' phenomenon between $S$ and $T_{[0,0]}$. 
To elucidate the above picture, we investigate the operator content of the two lowest singlets  by inspecting a traceless tensor: 
\begin{equation}
    T_{55}(\boldsymbol{r})=\int d\boldsymbol{r^\prime}\delta(\boldsymbol{r}-\boldsymbol{r^\prime})[n_5(\boldsymbol{r})n_5(\boldsymbol{r^\prime})-\frac{1}{5}\boldsymbol{n}(\boldsymbol{r})\boldsymbol{n}(\boldsymbol{r^\prime})].
\end{equation}
$T_{55}$ belongs to one of the 14 components of the SO(5) rank-2 tensor representation and it should capture the essence of $T_{[0,0]}$. In Fig. \ref{fig:overlap}(b), we find that $T_{55}$ dominates the lowest singlet when $0.1 \lesssim u_5<1$, while the second lowest singlet takes larger weight of $T_{55}$ when $u_5\lesssim 0.1$. 
Therefore, we conclude that operator content carried by the two lowest singlets exchange by varying $u_5$, producing an avoided level crossing behavior in Fig. \ref{fig:flow_even} and Fig. \ref{fig:overlap}(a). 
Physically, this is possible by considering the level of repulsive interaction between the primaries within the same representation group.
In the CFT formalism, the interaction between $S$ and $T_{[0,0]}$ is nonzero as long as the operator product expansion coefficient $f_{S S^{'} T_{[0,0]}}$ is finite ($S^{'}$ is a CFT operator). In brief, the observation of avoided level crossing in the RG flow is attributed to the interplay among primary states. 

\begin{figure}[t]
\includegraphics[width=0.48\linewidth]{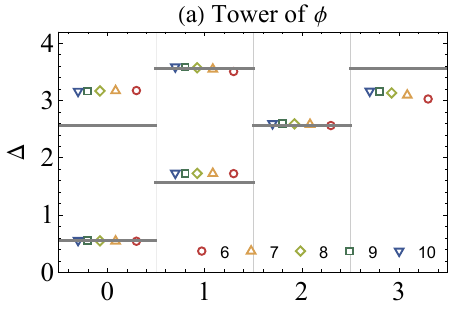}
\includegraphics[width=0.48\linewidth]{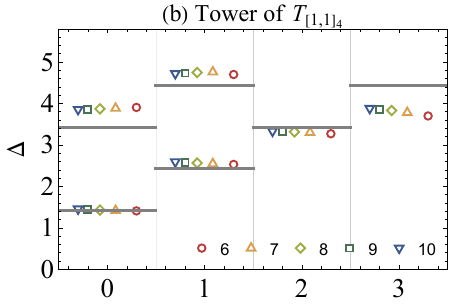}
\includegraphics[width=0.48\linewidth]{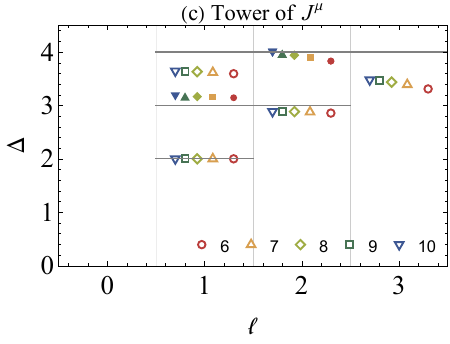}
\includegraphics[width=0.48\linewidth]{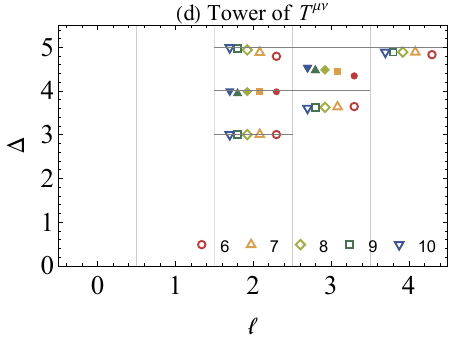}
	\caption{
    The operator spectra of the conformal multiplets for  (a) $\phi_{[\frac{1}{2},\frac{1}{2}]}$, (b) $T_{[1,1]}$, (c) $J^\mu$ and (d) $T^{\mu\nu}$. The horizontal grey lines denote the anticipated values from the conformal symmetry. These data are calculated at $u_5/u =- 0.75$. Various symbols stand for the system size with orbitals 6-10. 
    } 
	\label{fig:Tower}
\end{figure}

Since the operator drifts are slow for $u_5/u<-0.5$ in Fig. \ref{fig:flow_even}, we present 
the operator spectra as shown in Fig. \ref{fig:Tower}. 
Taking the vector primary $\phi_{[\frac{1}{2},\frac{1}{2}]}$,  rank-2 tensor primary $T_{[1,1]}$, current $J^{\mu}$ and energy-momentum tensor $T^{\mu\nu}$ as examples, we identify
all of their low-lying descendants with $\Delta \le 7$ and $\ell \le 4$. 
The scaling dimensions of descendants are very close to the integer spacing with the primary operator, supporting an approximate conformal symmetry that emerges at the O(4) fixed point. 
The discrepancy away from the exact conformal expectation is due to the appearance of relevant singlet primary $S$ (see below).

Even though the fixed point is not exactly hit since the conformal symmetry is not exact, the scaling dimensions of low-lying O(4) primaries drift very slowly within $u_5/u\in [-1,-0.5]$. Next, we present the estimate of scaling dimensions in Tab. \ref{tab:primary} at a representative parameter point $u_5/u=-0.75$.
1) The lowest O(4) parity-odd vector $\phi_{[\frac{1}{2},\frac{1}{2}]}$ corresponds to the order parameter, with $\Delta_{\phi_{[\frac{1}{2},\frac{1}{2}]}} \approx 0.555 \pm 0.010$, relating to the anomalous dimension $\eta =2(\Delta_{\phi_{[\frac{1}{2},\frac{1}{2}]}}-1/2)\sim 0.11 \pm0.02 $. 2) The lowest O(4) parity-even rank-2 tensor
$T_{[1,1]}$ corresponds to the relevant perturbation that controls the easy-plane N\'eel-to-VBS transition. Its scaling dimension $\Delta_{T} \sim 1.453 \pm 0.025$ is related
to the exponent $\nu=1/(3-\Delta_{T_{[1,1]}})\sim 0.65 \pm0.01$.
3) The $6\pi-$monopole operator is relevant, while higher monopole operators are irrelevant. It indicates that O(4) DQCP is not stable on the $C_3$-symmetric lattice model (e.g., honeycomb), which allows the $6\pi-$monopole operator. 
4) A parity-even scalar $S$ exists in the O(4) fixed point and it is relevant. Relevant $S$ suggests that the observed N\'eel-to-VBS transition in the O(4)-symmetric model is not captured by a genuine fixed point. Instead, the true fixed point lies outside the current parameter space but is very close to the parameter region we identified, allowing us to observe approximate conformal symmetry. 
The above observations are very similar to those in the SO(5) case \cite{zhou2024mathrmso5deconfinedphasetransition}.

\begin{table}[t!]
    \centering
    \caption{The scaling dimension and corresponding quantum numbers for the lowest lying primary operators obtained from state-operator correspondence on different system sizes $N_\mathrm{o}=7-10$. 
    Here, we choose $u_5/u=-0.75$ and set $\Delta_{{T}^{\mu\nu}}\approx 3,\Delta_{{J}^{\mu}}=2$. 
    }
    \setlength{\tabcolsep}{4.2pt}
    \begin{tabular}{c|ccc|cccc}
        \hline\hline
        \multicolumn{1}{c}{}&\multicolumn{3}{c|}{$N_\mathrm{o}$}&10&9&8&7\\
        \multicolumn{1}{c}{}&\multicolumn{3}{c|}{$u$}&$1.7959$&$1.1079$&$0.7885$&$0.6071$\\
        \hline
        Op.&$\ell$ & $\mathcal{P}$ & Rep. &&& $\Delta$ & \\\hline
        $\phi_{[\frac{1}{2},\frac{1}{2}]}$                  & $0$ & $+$ & ${[\frac{1}{2},\frac{1}{2}] }$ & $0.555$ & $0.550$ & $0.546$ & $0.542$  \\
        $T_{[1,1]}$                     & $0$ & $+$ & ${[1,1]}$ & $1.453$ & $1.438$ & $1.425$ & $1.413$  \\
        $J^\mu$                 & $1$ & $+$ & ${[0,1]}$ & $2.000$ & $2.000$ & $2.000$ & $2.000$ \\
        $\mathcal M_{6\pi}$     & $0$ & $-$ & ${[\frac{3}{2},\frac{3}{2}]}$ & $2.717$ & $2.610$ & $2.583$ & $2.559$  \\
        $S$                     & $0$ & $+$ & ${[0,0] }$ & $2.845$ & $2.840$ & $2.836$ & $2.833$  \\
        $\mathcal M_{8\pi}$     & $0$ & $+$ & ${[2,2]}$ & $4.069$ & $4.025$ & $3.981$ & $3.941$  \\
        \hline\hline
    \end{tabular}
    \label{tab:primary}
\end{table}

The above findings suggest a putative RG diagram as shown in Fig. \ref{fig:drawing}(b), i.e. 3D NLsM
with O(4) symmetry shares the similar properties with the SO(5) case \cite{zhou2024mathrmso5deconfinedphasetransition,Ma2020Theory,Nahum2020WZW}. The direct phase transitions between the Néel and VBS phases in the model equation \ref{eq:ham_real} fall into the pseudo-critical regime due to the relevance of scalar $S$.

\textit{Summary and discussion.---}
We have numerically studied the fate of the DQCP in a microscopic model with a tunable symmetry from SO(5) to O(4). Within the fuzzy sphere framework, we identify crucial renormalization group flow of conformal data.   
For example, O(4) primaries are decomposed from original SO(5) primaries. 
Especially the operator content of some operators with identical symmetries exchanges their operator content via the avoided level crossing. This phenomenon unveils intricate interactions among the operators. 
Moreover, the scalar operator $S$ persists relevantly from SO(5) to O(4), which inevitably leads to a weak first-order transition between the easy-plane N\'eel and VBS phases. 

A final remark concerns the field theory description of the O(4) DQCP, where its topology is encoded in a topological $\theta$-term \cite{ChongWang-PhysRevX.7.031051}. This implies that the local operator $\sim \epsilon^{\alpha\beta\gamma\delta}\phi_\alpha \partial_t \phi_\beta \partial_x \phi_\gamma \partial_y \phi_\delta$ is expected to be relevant at the O(4) DQCP. In our symmetry analysis, this operator should be a parity-odd scalar. However, our extensive calculations revealed no additional relevant parity-odd scalars within the spectrum of operators. Resolving the absence of this theoretically expected operator remains a significant question for future investigation.

Furthermore, 
to investigate the RG flows of conformal operators is challenging for traditional methods. Here, we point out that the fuzzy sphere regularization scheme offers a new perspective for inspecting the RG flow process in a microscopic way.  
We envision that this scheme could advance future studies of RG flows among various fixed points in critical phenomena (e.g. another example of Wilson-Fisher O(3)$\rightarrow$O(2) flow is shown in Supple. Mat. \cite{SM}).


\begin{acknowledgments}
We thank Zheng Zhou, Chong Wang, Jie Lou, and Zhenjiu Wang for the fruitful discussion. The numerical calculation partially relies on the FuzzifiED\cite{zhou2025fuzzifiedjuliapackage} libraries. S. Y. and Y. C. are supported by the National Key Research and Development Program of China Grant No. 2022YFA1402204 and the National Natural Science Foundation of China Grant No. 12274086. This work was also supported by NSFC No. 12474144 (L.D.H, C.H., W.Z.).

\end{acknowledgments}

	

\bibliography{O4Sphere.bib}


\clearpage
\begin{widetext}
\startsupplement
\begin{center}
\textbf{Supplemental Material}    
\end{center}

This Supplemental Material includes: 1) The analysis of branching rules from SO(5) to SO(4) groups(Sec I). 2)The second-quantized Hamiltonian and several discrete symmetries can be utilized to accelerate numerical calculations(Sec II). 3) The distribution of excitation energy gaps in the model under different u5 anisotropy strengths (Sec III). 4) The operator flow behavior from the O(3)-Heisenberg fixed point to the O(2)-XY fixed point(Sec IV).
\section{Analysis of O(4) Symmetry}
\subsection{$\mathfrak{so}(4)$ Lie algebra}
The generators of $\mathfrak{so}(5)$ Lie algebra are
\begin{equation}
    \Gamma^{ab} = -\frac{i}{2}[\Gamma^a, \Gamma^b]=-i\Gamma^a\Gamma^b
\end{equation}
where $\Gamma^a$ are Gamma matrices
\begin{equation}
    \Gamma^1 = \begin{pmatrix}
        0& I\\ I&0
    \end{pmatrix}\quad
    \Gamma^2 = \begin{pmatrix}
        0& -iI\\ iI&0
    \end{pmatrix}\quad
    \Gamma^{3-5} = \begin{pmatrix}
        \vec{\sigma}& 0\\ 0&-\vec{\sigma}
    \end{pmatrix}
\end{equation}
The commutation relations of the $ \mathfrak{so}(5) $ Lie algebra are
\begin{equation}
    [\Gamma^{ab}, \Gamma^{cd}] = 2i\{ \delta^{ab}\Gamma^{bd}+ \delta^{bd}\Gamma^{ac}-\delta^{ad}\Gamma^{bc}-\delta^{bc}\Gamma^{ad}\}
\end{equation}
If we remove \( \Gamma^5 \), the remaining 6 \( \mathfrak{so}(5) \) generators form exactly an \( \mathfrak{so}(4) \) Lie algebra. Since \( \mathfrak{so}(4) = \mathfrak{su}(2) \oplus \mathfrak{su}(2) \), we will proceed by constructing the generators of these two \( \mathfrak{su}(2) \) algebras.

The first $\mathfrak{su}(2)$ sub-algebra is formed by 
\begin{equation}
    \sigma^z_1 = \frac{\Gamma^{12}+\Gamma^{34}}{2} = \begin{pmatrix}
        1&0&0&0\\0&0&0&0\\0&0&0&0\\0&0&0&-1
    \end{pmatrix}\qquad
    \sigma^x_1 = \frac{\Gamma^{23}+\Gamma^{14}}{2} = \begin{pmatrix}
        0&0&0&1\\0&0&0&0\\0&0&0&0\\1&0&0&0
    \end{pmatrix}\qquad
    \sigma^y_1 = \frac{\Gamma^{24}-\Gamma^{13}}{2} = \begin{pmatrix}
        0&0&0&-i\\0&0&0&0\\0&0&0&0\\i&0&0&0
    \end{pmatrix}.
\end{equation}
The three generators above form an \( \mathfrak{su}(2) \) Lie algebra, and the Casimir operator is 
\begin{equation}
    \bm{s}_1^2 = (s_1^x)^2+(s_1^y)^2+(s_1^z)^2
\end{equation}
where $s^{x,y,z} = \frac{\hbar}{2}\sigma^{x,y,z}$.

The second $\mathfrak{su}(2)$ sub-algebra is formed by
\begin{equation}
    \sigma^z_2 = \frac{\Gamma^{12}-\Gamma^{34}}{2} = \begin{pmatrix}
        0&0&0&0\\0&1&0&0\\0&0&-1&0\\0&0&0&0
    \end{pmatrix}\qquad
    \sigma^x_2 = \frac{\Gamma^{23}-\Gamma^{14}}{2} = \begin{pmatrix}
        0&0&0&0\\0&0&1&0\\0&1&0&0\\0&0&0&0
    \end{pmatrix}\qquad
    \sigma^y_2 = \frac{\Gamma^{24}+\Gamma^{13}}{2} = \begin{pmatrix}
        0&0&0&0\\0&0&-i&0\\0&i&0&0\\0&0&0&0
    \end{pmatrix}.
\end{equation}
The Casimir operator is 
\begin{equation}
    \bm{s}_2^2 = (s_2^x)^2+(s_2^y)^2+(s_2^z)^2.
\end{equation}

Since \( \mathfrak{so}(4) = \mathfrak{su}(2) \oplus \mathfrak{su}(2) \), representations of \( \mathfrak{so}(4) \) can be constructed from the representations of these two \( \mathfrak{su}(2) \) algebras. Let \( s_1 \) and \( s_2 \) denote the highest weights of the two \( \mathfrak{su}(2) \) representations, which means that the eigenvalues of the two Casimir operators are \( (s_1 + 1)s_1 \) and \( (s_2 + 1)s_2 \), respectively. We denote this representation as \( [s_1, s_2] \). This representation contains a total degeneracy of \( (2s_1 + 1)(2s_2 + 1) \), with values ranging as \( -s_1 \leq s_1^z \leq s_1 \) and \( -s_2 \leq s_2^z \leq s_2 \), and includes no internal degeneracies. The only degeneracy arises because the two \( \mathfrak{su}(2) \) algebras are equivalent, so the \( [s_1, s_2] \) representation has the same energy as the \( [s_2, s_1] \) representation. In Tab. \ref{tbl:rep_deg}, we listed the degeneracy of each representation in sector $(s_1^z, s_2^z)$.

\begin{table}[!htb]
    \centering
    \setlength{\tabcolsep}{7pt}
    \caption{The quadratic Casimir $C_2$ of different $\mathfrak{so}(4)=\mathfrak{su}(2) \oplus \mathfrak{su}(2)$ representations and the corresponding state degeneracies in different $(s^z_1,s^z_2)$  sectors. Here, we only listed the sectors with $0\le s^z_2\le s^z_1$. The single number $1,2,3\cdots$ means sectors $(\pm s^z_1,\pm s^z_2)$ should have the same degeneracy $1,2,3\cdots$, and the subscript $~_2$ in $1_2,2_2,3_2\cdots$ means the sectors $(\pm s^z_1,\pm s^z_2)$ and $(\pm s^z_2,\pm s^z_1)$ should have the same degeneracy $1,2,3\cdots$.}
    \begin{tabular}{cl|ccccccccccc}
        \hline\hline
        Rep.&$\mathrm{dim}$&\multicolumn{11}{c}{Degeneracy in sector $(s^z_1,s^z_2)$}\\
        $[s_1,s_2]$& &$(0,0)$&$(1,0)$&$(1,1)$&$(2,0)$&$(2,1)$&$(2,2)$&$(3,0)$&$(3,1)$&$(3,2)$&$(3,3)$&$\cdots$\\
        \hline
        $[0,0]  $& 1 &1& & & & & & & & & &$\cdots$ \\
        $[1,0]  $& 6 &2&$1_2$& & & & & & & & &$\cdots$ \\
        $[1,1]  $& 9 &1&$1_2$&1& & & & & & & &$\cdots$ \\
        $[2,0]  $& 10&2&$1_2$& &$1_2$& & & & & & &$\cdots$ \\
        $[2,1]  $& 30&2&$2_2$&2&$1_2$&$1_2$& & & & & &$\cdots$ \\
        $[2,2]  $& 25&1&$1_2$&1&$1_2$&$1_2$&1& & & & &$\cdots$ \\
        $[3,0]  $& 14&2&$1_2$& &$1_2$& & &$1_2$& & & &$\cdots$ \\
        $[3,1]  $& 42&2&$2_2$&2&$1_2$&$1_2$& &$1_2$&$1_2$& & &$\cdots$ \\
        $[3,2]  $& 70&2&$2_2$&2&$2_2$&$2_2$&2&$1_2$&$1_2$&$1_2$& &$\cdots$ \\
        $[3,3]  $& 49&1&$1_2$&1&$1_2$&$1_2$&1&$1_2$&$1_2$&$1_2$&1&$\cdots$ \\
        \hline\hline
    \end{tabular}
    \begin{tabular}{cl|ccccccccccc}
        \hline\hline
        Rep.&$\mathrm{dim}$&\multicolumn{11}{c}{Degeneracy in sector $(s^z_1,s^z_2)$}\\
        $[s_1,s_2]$& &$(\frac12,\frac12)$&$(\frac32,\frac12)$&$(\frac32,\frac32)$&$(\frac52,\frac12)$&$(\frac52,\frac32)$&$(\frac52,\frac52)$&$(\frac72,\frac12)$&$(\frac72,\frac32)$&$(\frac72,\frac52)$&$(\frac72,\frac72)$&$\cdots$\\
        \hline
        $[\frac12,\frac12]  $& 4 &1& & & & & & & & & &$\cdots$ \\
        $[\frac32,\frac12]  $& 16 &2&$1_2$& & & & & & & & &$\cdots$ \\
        $[\frac32,\frac32]  $& 16 &1&$1_2$&1& & & & & & & &$\cdots$ \\
        $[\frac52,\frac12]  $& 24 &2&$1_2$& &$1_2$& & & & & & &$\cdots$ \\
        $[\frac52,\frac32]  $& 48 &2&$2_2$&2&$1_2$&$1_2$& & & & & &$\cdots$ \\
        $[\frac52,\frac52]  $& 36 &1&$1_2$&1&$1_2$&$1_2$&1& & & & &$\cdots$ \\
        $[\frac72,\frac12]  $& 32 &2&$1_2$& &$1_2$& & &$1_2$& & & &$\cdots$ \\
        $[\frac72,\frac32]  $& 64 &2&$2_2$&2&$1_2$&$1_2$& &$1_2$&$1_2$& & &$\cdots$ \\
        $[\frac72,\frac52]  $& 96 &2&$2_2$&2&$2_2$&$2_2$&2&$1_2$&$1_2$&$1_2$ & &$\cdots$ \\
        $[\frac72,\frac72]  $& 64 &1&$1_2$&1&$1_2$&$1_2$&1&$1_2$&$1_2$&$1_2$&1&$\cdots$ \\
        \hline\hline
    \end{tabular}
    \label{tbl:rep_deg}
\end{table}

\subsection{Branch rule of $\mathfrak{so}(5)\supset\mathfrak{so}(4)$}
In previous work on SO(5) DQCP\cite{zhou2024mathrmso5deconfinedphasetransition}, numerical calculations can be simplified using branching rules. In this subsection, we also list the branching rules for $\mathfrak{so}(5)\supset\mathfrak{so}(4)$, which can be used to analyze how the fields of the SO(5) DQCP evolve into the fields of O(4) case when the system's symmetry is explicitly broken from SO(5) to O(4).  Each irreducible representation of SO(5) is labeled by two non-negative integers $(p \ge q)$, and the corresponding Casimir and representation dimension are given by\cite{SO5LL}
\begin{equation}
\begin{split}
    \lambda(p,q) &= \frac14 p^2 +\frac14 q^2+p+\frac12 q\\
    D(p,q) &= \frac16 (p+2)(q+1)(p+q+3)(p-q+1)
\end{split}
\end{equation}
and the branch rule
\begin{equation}
    [p,q]_5 = \underset{0\le n \le q}{\oplus}
    \underset{-\frac{p-q}{2}\le s \le \frac{p-q}{2}}{\oplus} [j,k]_4
\end{equation}
where
\begin{equation}
    [j,k]_4 = [\frac{n}{2}+\frac{p-q}{4}+\frac{s}{2},\frac{n}{2}+\frac{p-q}{4}-\frac{s}{2}]_4.
\end{equation}
The subscripts 4 and 5 represent the irreducible representations of SO(4) and SO(5), respectively, while $j$ and $k$ denote the irreducible representations of the two SU(2) subgroups within SO(4). In the Tab.\ref{tbl:branch}, we present the lower-dimensional irreducible representations of SO(5) as examples.

Here, we would like to analyze the symmetry of primaries by breaking the symmetry from SO(5) to O(4). 
\begin{enumerate}
    \item The SO(5) order parameter $\phi_{[1,1]_5} \sim(n_1,n_2,n_3,n_4,n_5)$ transforms under the representation $[1,1]_5$ and has five components.  When the symmetry is broken from SO(5) to O(4), this primary field splits into two distinct fields, $\phi_{[\frac{1}{2}, \frac{1}{2}]_4}\sim(n_1,n_2,n_3,n_4)$ and $\phi^-_{[0,0]_4}\sim n_5$, corresponding to the representations $[\frac{1}{2}, \frac{1}{2}]_4$ and $[0,0]_4$, respectively. The former, $\phi_{[\frac{1}{2}, \frac{1}{2}]_4}$, serves as the order parameter for the O(4) symmetry breaking, while the latter, $\phi^-_{[0,0]_4}$ is a parity-odd field.
    \item The SO(5) rank-2 tensor $T_{ab} \sim n_a n_b - \delta_{ab} n^2 / 5$ (where $a, b = 1, 2, 3, 4, 5$) is the field that governs the Neel-VBS phase transition and has 14 independent components. When the symmetry is broken to O(4), $T_{ab}$ splits into three fields: $ (T_{[1,1]_4})_{ab} \sim n_a n_b - \delta_{ab} n^2 / 4$ (where $a, b = 1, 2, 3, 4 $), $ ((T_{[\frac{1}{2}, \frac{1}{2}]_4}))_a \sim n_a n_5 $, and $T_{[0,0]_4} \sim n_5 n_5$, corresponding to the representations $[1,1]_4$, $[\frac{1}{2}, \frac{1}{2}]_4$, and $[0,0]_4$, respectively. Among these, $ T_{[1,1]_4} $ and $T_{[0,0]_4}$ are crucial for studying the O(4) critical point. The $ (T_{[1,1]_4})_{ab} $ corresponds to the field governing the Neel-VBS phase transition, while the $T_{[0,0]_4}$ is a parity-even scalar.
    \item In SO(5), the $6\pi$-monopole operator $\mathcal{M}_{6\pi}$ is a rank-3 tensor in the $[3,3]_5$ representation. In SO(4) theory, it splits into four distinct operators:  
a) the $6\pi$-monopole operator $\mathcal{M}_{6\pi~[3/2,3/2]_4}$,  
b) a parity-odd operator $\mathcal{M}^-_{6\pi~[1,1]_4}$ sharing the same representation as $(T_{[1,1]_4})_{ab}$,  
c) a parity-even vector operator $\mathcal{M}_{6\pi~[1/2,1/2]_4}$,  
d) and a parity-odd scalar operator $\mathcal{M}^-_{6\pi~[0,0]_4}$.
\end{enumerate}

\begin{table}[!htb]
    \centering
    \setlength{\tabcolsep}{7pt}
    \caption{ The branch rule of $\mathfrak{so}(5)\supset\mathfrak{so}(4)$. The subscripts 4 and 5 represent the irreducible representations of SO(4) and SO(5), respectively. }
    \begin{tabular}{ccc|c}
        \hline\hline
        $[p,q]_5$&$\text{Dim}_{SO(5)}$&$\text{Casimir}_{SO(5)}$&Branch rule of $\mathfrak{so}(5)\supset\mathfrak{so}(4)$\\
        \hline
        $[0,0]_5$& 1 &0& $[0,0]_4$  \\
        $[1,1]_5$& 5 &2& $[0,0]_4\oplus[\frac12,\frac12]_4$  \\
        $[2,0]_5$& 10 &3& $[1,0]_4\oplus[\frac12,\frac12]_4\oplus[0,1]_4$  \\
        $[2,2]_5$& 14 &5& $[0,0]_4\oplus[\frac12,\frac12]_4\oplus[1,1]_4$  \\
        $[3,1]_5$& 35 &6& $[0,1]_4\oplus[\frac12,\frac12]_4\oplus[1,0]_4\oplus[\frac12,\frac32]_4\oplus[1,1]_4\oplus[\frac32,\frac12]_4$  \\
        $[3,3]_5$& 30 &9& $[0,0]_4\oplus[\frac12,\frac12]_4\oplus[1,1]_4\oplus[\frac32,\frac32]_4$  \\
        $[4,0]_5$& 35 &8& $[0,2]_4\oplus[\frac12,\frac32]_4\oplus[1,1]_4\oplus[\frac32,\frac12]_4\oplus[2,0]_4$  \\
        \hline\hline
    \end{tabular}
    \label{tbl:branch}
\end{table}

\section{Second quantization form of Hamiltonian}

By projecting the model Hamiltonian into the LLL, we obtain the second quantization form of Hamiltonian in orbital space reads \cite{ZHHHH2022}
\begin{small}
    \begin{equation}
\begin{split}
H_{\rm{LLL}} = &\sum_{m_1m_2m_3m_4} V_{m_1m_2m_3m_4} \bigg[  \boldsymbol{c}_{m_1}^\dag \boldsymbol{c}_{m_4} \boldsymbol{c}_{m_2}^\dag \boldsymbol{c}_{m_3} \\
& - \sum_{i=1}^5 u_i \boldsymbol{c}_{m_1}^\dag \Gamma^i \boldsymbol{c}_{m_4} \boldsymbol{c}_{m_2}^\dag \Gamma^i \boldsymbol{c}_{m_3} \bigg]\delta_{m_1+m_2,m_3+m_4},
\end{split}
\label{eq:hamLLL}
\end{equation}
\end{small}
where the quantity $V_{m_1m_2m_3m_4}$ is interaction element of  
the short-ranged potential $\delta(\boldsymbol{r}_{1}-\boldsymbol{r}_{2})$ between fermions in real space:
\begin{small}
    \begin{equation}
V_{{m_1m_2m_3m_4}}=
\sum_lV_l(4s-2l+1)\begin{pmatrix}
s&s&2s-l\\m_1&m_2&-m_1-m_2
\end{pmatrix}\begin{pmatrix}
s&s&2s-l\\m_4&m_3&-m_4-m_3
\end{pmatrix},
\end{equation}
\end{small}
Throughout the paper, we  consider fermions with half-filling of the LLL. 

This model has several discrete symmetries. We list some useful discrete symmetries that can be utilized to simplify numerical calculations.
\paragraph{Valley $\mathbb{Z}_2$}
The valley \( \mathbb{Z}_2 \) operator exchanges the two \( \mathfrak{su}(2) \) subalgebras:
\begin{equation}
    \mathbb{Z}_2^{\text{valley}} = \begin{pmatrix}
        0&1&0&0\\1&0&0&0\\0&0&0&1\\0&0&1&0
    \end{pmatrix}
\end{equation}

\paragraph{PH-symmetry}
The PH-symmetry(or parity) is defined as
\begin{equation}
    \mathcal{P}: \begin{pmatrix}
        \hat c_1\\\hat c_2\\\hat c_3\\\hat c_4
    \end{pmatrix}\rightarrow
    \begin{pmatrix}
        0&0&0&-1\\0&0&-1&0\\0&1&0&0\\1&0&0&0
    \end{pmatrix}\begin{pmatrix}
        \hat c_1^\dagger\\\hat c_2^\dagger\\\hat c_3^\dagger\\\hat c_4^\dagger
    \end{pmatrix}
\end{equation}

\paragraph{Spin $\mathbb{Z}_2$}
The spin \( \mathbb{Z}_2 \) operator exchanges spin up and down in each valley:
\begin{equation}
\begin{split}
    \mathbb{Z}_2^{\text{spin}}(1) &= \begin{pmatrix}
        0&0&0&1\\0&0&0&0\\0&0&0&0\\1&0&0&0
    \end{pmatrix}\\
    \mathbb{Z}_2^{\text{spin}}(2) &= \begin{pmatrix}
        0&0&0&0\\0&0&1&0\\0&1&0&0\\0&0&0&0
    \end{pmatrix}
\end{split}
\end{equation}

\paragraph{$\pi$ rotation around $y$ axis} The $\pi$ rotation around $y$ axis is available only in sector $L_z=0$:

\begin{equation}
    \mathcal{R}_y: \hat{\bm{c}}_m \rightarrow \hat{\bm{c}}_{-m}
\end{equation}

\section{evolution of excited gaps with $u_5$ anisotropy}
The Fig. \ref{fig:SO5_O4_gap_with_u5} shows the evolution of the bare excitation energy gaps (calculated at finite size $N_{\mathrm{o}}=9$) as a function of the anisotropy strength $u_5$. The point $u_5/u = 1$ corresponds to the SO(5)-symmetric point. When $u_5/u > 1$, the system favors polarized along the direction of the 5-th vector, and the ground state develops a finite expectation value of $ \langle n_5 \rangle$. Depending on the sign of $\langle n_5 \rangle$, the ground state should be two-fold degenerate in the thermodynamic limit. For $u_5/u < 1$, the symmetry of the model is reduced to O(4). Take the lowest excitation $\phi$ as an example: at the SO(5) symmetric point, it should be 5-fold degenerate, while under O(4) symmetry, it splits into a 4-fold degenerate $ \phi_{[\frac{1}{2},\frac{1}{2}]}$. The remaining state $\phi_{[0,0]}^{-}$ transforms as the [0,0] representation and is expected to be a gapped state (as analyzed in the main text).
\begin{figure}
    \centering
    \includegraphics[width=0.5\linewidth]{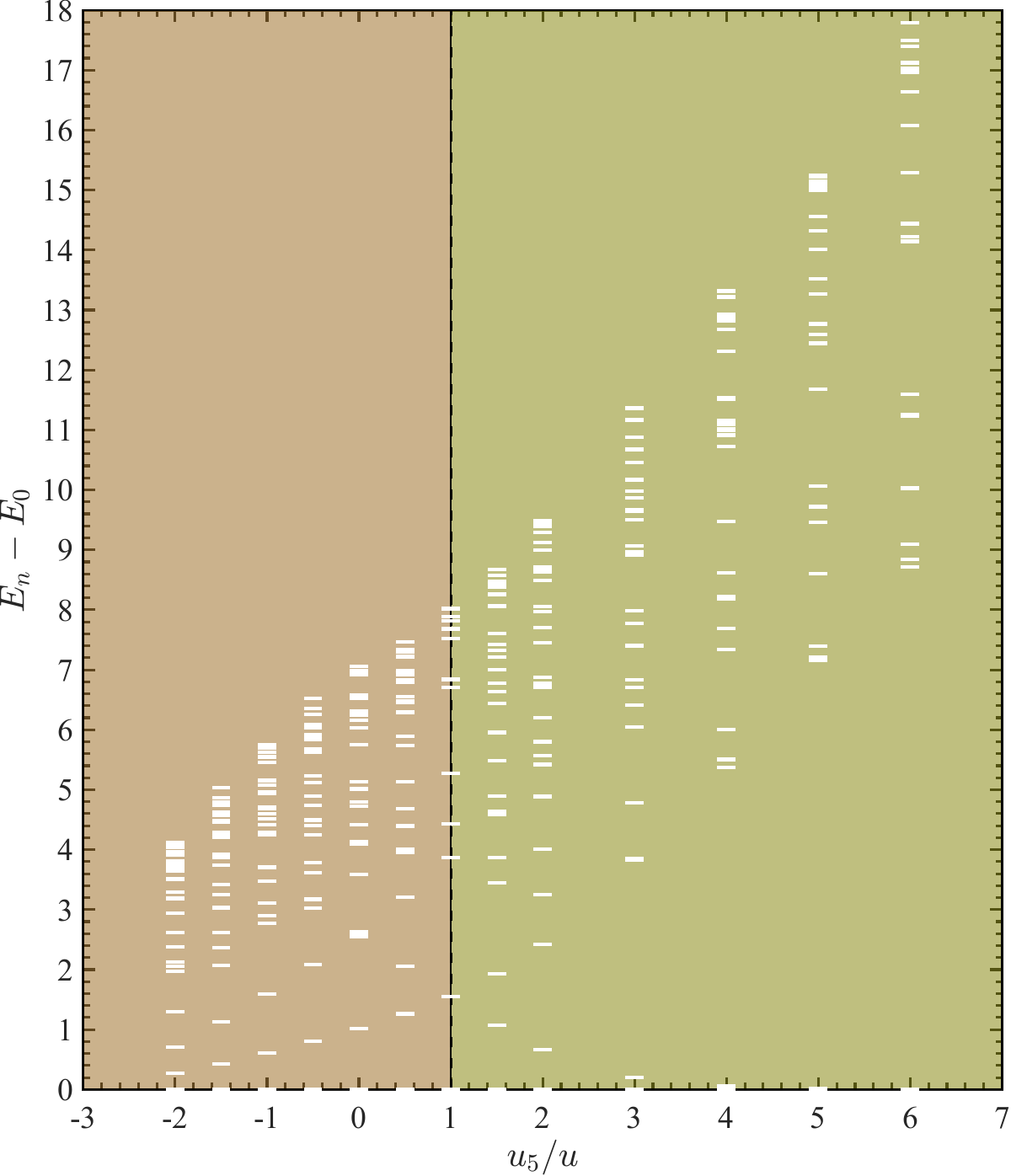}
    \caption{The lowest 50 excited gaps in the $L_z=0$ sector as a function of $u_5$, in the model of Neel-VBS transition (Eq. 1 in the main text). }
    \label{fig:SO5_O4_gap_with_u5}
\end{figure}

\section{Spectral flow from SO(5) to O(4)}
In Fig. \ref{fig:flow}, we observe that primaries with representation under the O(4) group are decomposed from original SO(5) primaries. For example, the SO(5) primary $\phi$ splits into a O(4)  vector $\phi_{[\frac{1}{2},\frac{1}{2}]}$ and a O(4) parity-odd scalar $\phi^-_{[0,0]}$; the SO(5) current $J^\mu$ splits into a O(4) current $J^{\mu}_{[1,0]}(J^{\mu}_{[0,1]}$) and a parity-odd vector $J^{\mu,-}_{[\frac{1}{2},\frac{1}{2}]}$.   
Here $'-'$ means parity-odd. 
These findings are consistent with the symmetry analysis in Sec. I.B. and in Tab. \ref{tbl:branch}.

By inspecting these flows in detail, we identify the splitting parity-odd  operators evolve very quickly, such as $\phi^-_{[0,0]}, T^-_{[\frac{1}{2},\frac{1}{2}]}$ as shown by the dashed lines in Fig. \ref{fig:flow}. A further extrapolation of their absolute energy gap indicates that these operators are gapped, corresponding to non-conformal fields (Fig. \ref{fig:gap}). Therefore, we only show the spectral flow of parity-even operators, as shown in Fig. 2 of the main text. 

\begin{figure}[b]
\includegraphics[width=0.97\linewidth]{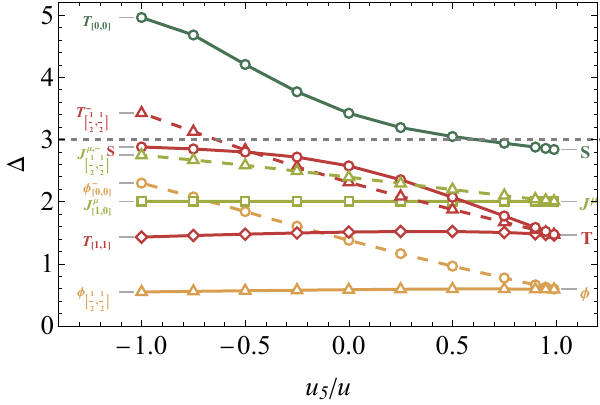}
	\caption{ 
    Evolution of low-lying operators with varying $u_5/u$. The SO(5) vector $\phi$ splits into the O(4) vector $\phi_{[\frac{1}{2},\frac{1}{2}]}$ and a $O(4)$ parity-add scalar $\phi^-_{[0,0]}$; The SO(5) current $J^\mu$ splits into a O(4) current $J^\mu_{[1,0]}(J^\mu_{[1,0]})$ and parity-odd vector $J^{\mu,-}_{[\frac{1}{2},\frac{1}{2}]}$; The SO(5) rank-2 tensor $T$ splits into  $T_{[1,1]}$, $T_{[0,0]}$, $T^{-}_{[\frac{1}{2},\frac{1}{2}]}$. Here, $[j,k]$ denotes the irreducible representations of two SU(2) subgroups within SO(4) and `$-$' means parity-odd.  
    } 
	\label{fig:flow}
\end{figure}

\begin{figure}[t]
\includegraphics[width=0.47\linewidth]{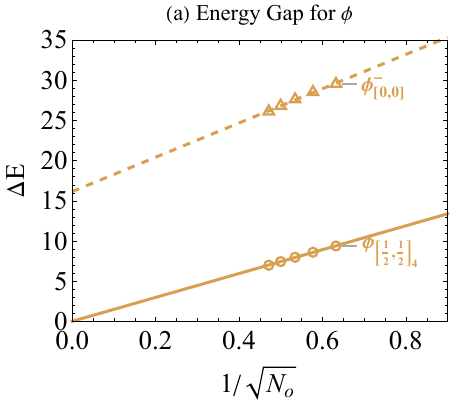}
\includegraphics[width=0.47\linewidth]{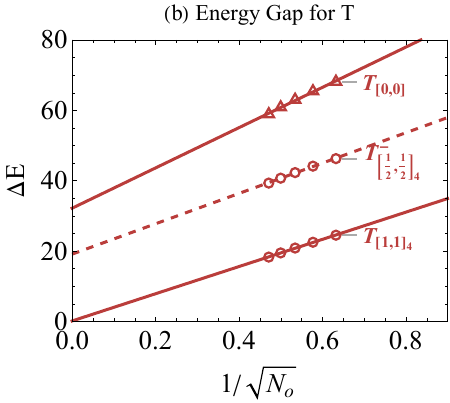}
	\caption{ The energy gap $\Delta E_{\mathcal O}=E_{\mathcal O}-E_0$ for (a) $\phi_{[\frac12,\frac12]}$ and $\phi^-_{[0,0]}$ split from the SO(5) vector primary $\phi$, and (b) $T_{[1,1]}$, $T^-_{[\frac12,\frac12]}$ and $T_{[0,0]}$ split from the SO(5) rank-2 primary $T$. Lines are finite-size extrapolations of the energy gap as a function of $1/\sqrt{N_{\text{o}}}$, showing that only fields $\phi_{[\frac12,\frac12]}$ and $T_{[1,1]}$ are gapless in the thermodynamic limit. } 
	\label{fig:gap}
\end{figure}

\section{operator flows from wilson-fisher O(3) to O(2) fixed point}
As supporting evidence, we also examined the flow from the well-understood Wilson-Fisher O(3) fixed point to an O(2) fixed point, for which relatively precise conformal data are available. In tracking the operator flow from the  O(3) to the O(2) fixed point, the key questions to be addressed in this section are:  
1) How do the primary operators of O(3) evolve into various operators under O(2)?  
2) Can similar level crossing behavior be observed?  
3) Compared to the operator flow from SO(5) to  O(4), what are the differences? Specifically, we considered a model that realizes the O(3) Heisenberg universality class\cite{han2023conformaloperatorcontentwilsonfisher}:
\begin{small}
        \begin{align}
	H_\textrm{O(3)}=\! \int\!d\boldsymbol{r}_{1}d\boldsymbol{r}_2\! \left[ \boldsymbol{n}(\boldsymbol{r}_1)\boldsymbol{n}(\boldsymbol{r}_2)+ U_2 \boldsymbol{n}_1(\boldsymbol{r}_1) \cdot \boldsymbol{n}_2(\boldsymbol{r}_2) 
	- U_1 (\boldsymbol{n}_1(\boldsymbol{r}_1) \cdot \boldsymbol{n}_1(\boldsymbol{r}_2) + \boldsymbol{n}_2(\boldsymbol{r}_1) \cdot \boldsymbol{n}_2(\boldsymbol{r}_2)) 
	  \right]  \!
	- h \int d\boldsymbol{r} \hat{\mathbf{\Psi}}^\dagger \tau^x\sigma^0 \hat{\mathbf{\Psi}},
    \label{eq:O3_ham}
\end{align}
\end{small}
The 4-flavor fermion $\boldsymbol{\Psi}$ accounts for both the layer degree of freedom $\tau = 1, 2$ and the spin degree of freedom $\sigma = \uparrow, \downarrow$. To realize O(2)-XY phase transition, we need to introduce an easy-plane anisotropy term:
\begin{small}
    \begin{equation}
    H_{zz}=\int d{\boldsymbol{r}}_{1}d\boldsymbol{r}_2 U_2(\Delta_z-1)[n_1^z(\boldsymbol{r}_1)n_2^z(\boldsymbol{r}_2)-\frac{1}{3}{\boldsymbol{n}}_1(\boldsymbol{r}_1)\cdot{\boldsymbol{n}}_2(\boldsymbol{r}_2)]=\int d{\boldsymbol{r}_{1}}d\boldsymbol{r}_2U_2[\frac{3-r_z}{2}(n_1^xn_2^x+n_1^yn_2^y)+r_zn_1^zn_2^z],\text{let } r_z=\frac{1+2\Delta_z}{3}
\end{equation}
\end{small}
Thus, the symmetry of the model $H_{\text{O(2)}}=H_{\text{O(3)}}+H_{zz}$ can be reduced from O(3) to O(2) by tuning parameter $r_z$. 
Fig. \ref{fig:gaps_along_O3_O2_flow} shows the evolution of excitation gaps as a function of $r_z$. When $r_z=1$, the model hosts O(3)-Heisenberg phase transition, the parameter setting of O(3) transition point is identical with Ref. \cite{han2023conformaloperatorcontentwilsonfisher}.  When $r_z > 1$, the system tends to polarize along the $z$-component direction (easy-axis case). The two lowest energy states become nearly degenerate, while the other states are nearly gapped. When $r_z < 1$, the system possesses O(2) symmetry (easy-plane case). Under various easy-plane anisotropy strengths $r_z$, the transverse field $h$ can also drive a phase transition from an O(2) symmetry-breaking magnetic ordered XY phase into an O(2)-symmetric disordered paramagnetic phase.
\begin{figure}
    \centering
    \includegraphics[width=0.5\linewidth]{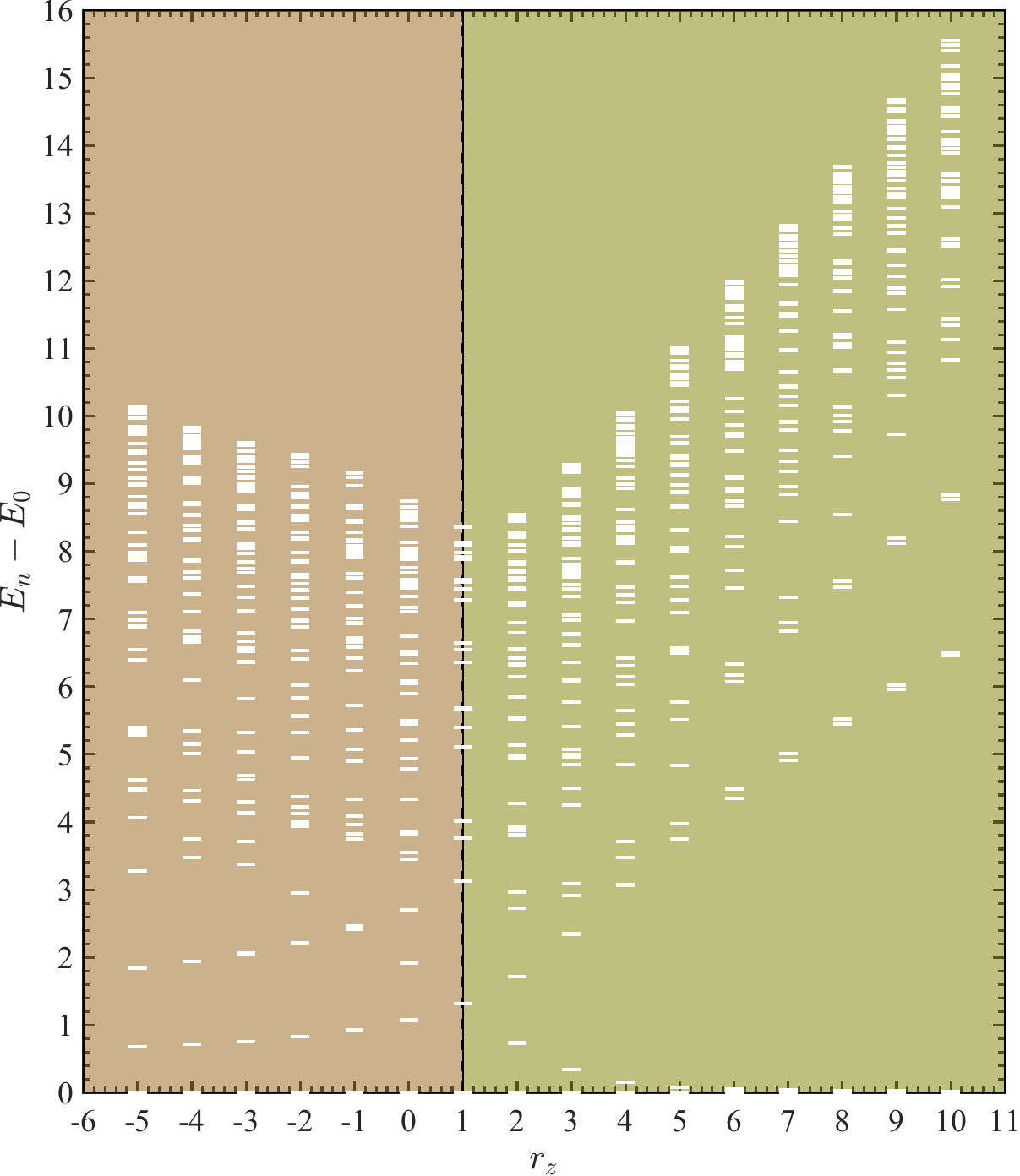}
    \caption{The lowest 50 excited gaps in the $L_z=0$ sector as a function of $r_z$, in the model of $H_{\text{O(2)}}=H_{\text{O(3)}}+H_{zz}$. }
    \label{fig:gaps_along_O3_O2_flow}
\end{figure}

The phase transition can be characterized by the following order parameter. 
\begin{equation}
    \boldsymbol{M}_{\tau,{\text{XY}}}=\sum_{m,\tau} \boldsymbol{c}_{m,\tau}^\dagger \boldsymbol{\sigma}_{xy}\boldsymbol{c}_{m,\tau}
\end{equation}
Here, $\boldsymbol{\sigma}_{xy}=\{\sigma_x,\sigma_y\}$. Then
we use the crossing points of the Binder ratio $U_4=\left<\boldsymbol{M}_{\tau,{\text{XY}}}^4\right>/\left<\boldsymbol{M}_{\tau,{\text{XY}}}^2\right>^2$ computed for different system sizes to extrapolate the critical value $h_c$ as illustrated in Fig. \ref{fig:XY_transition}.
\begin{figure}
    \centering
    \includegraphics[width=0.45\linewidth]{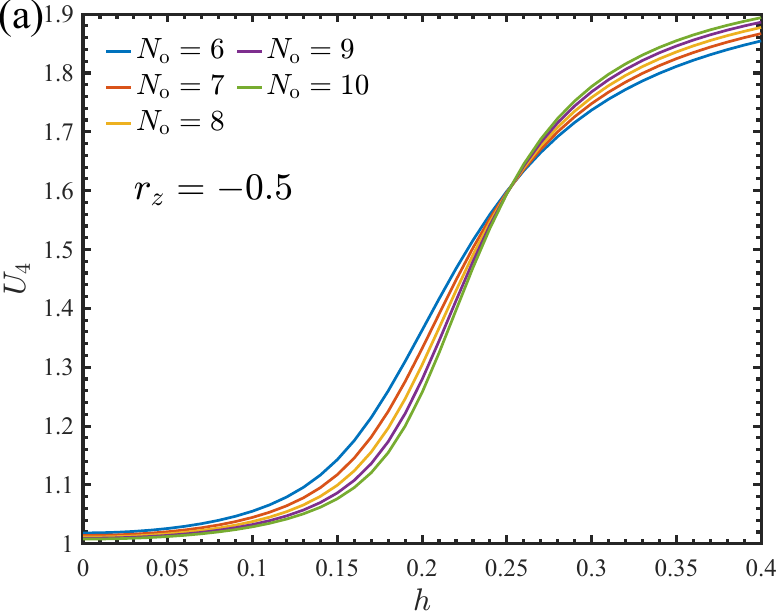}
    \includegraphics[width=0.45\linewidth]{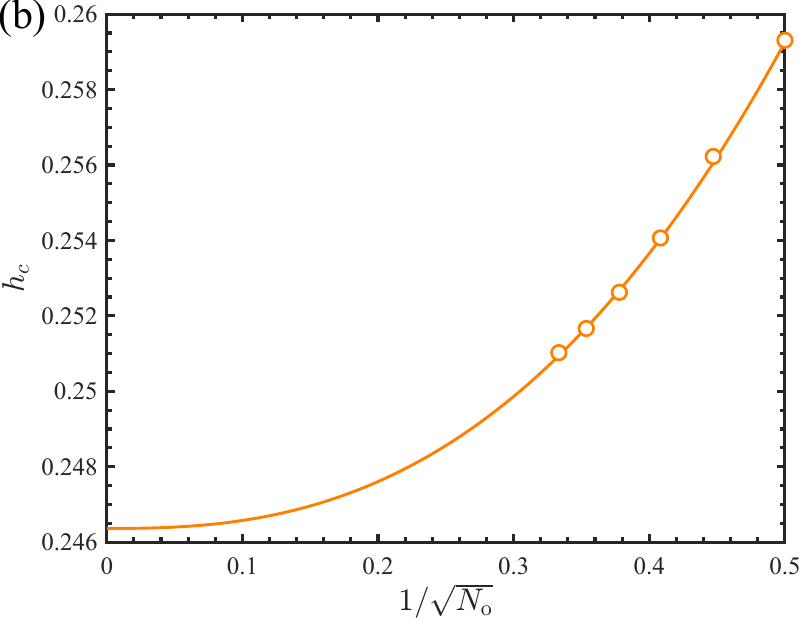}
    \caption{(a) Binder ratio $U_4$ versus transverse field $h$, data from different system sizes intersect nicely at the same point.  
(b) Using the extrapolation formula $h_c = h_c(\infty) + a N_{\text{o}}^{-b/2}$, we obtain the critical point $h_c(\infty)\approx0.2463$.}
    \label{fig:XY_transition}
\end{figure}

At these critical points, we compute the energy spectrum and extract the scaling dimensions of operators using the state-operator correspondence. Fig. \ref{fig:O3_O2_Op_flow} shows how the primary fields of the O(3) theory evolve into the primary fields of the O(2) theory.
\begin{figure}
    \centering
    \includegraphics[width=0.5\linewidth]{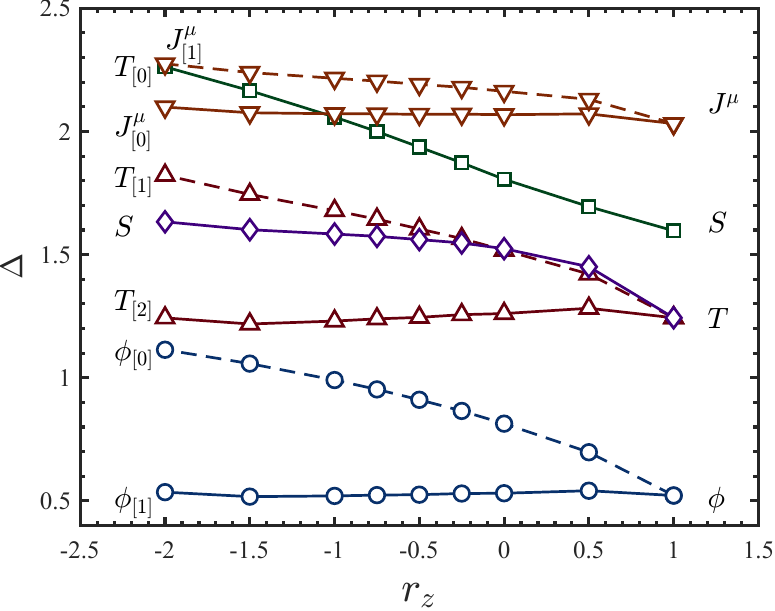}
    \caption{Evolution of several lowest primary operators along the flow from O(3) to O(2) fixed point.}
    \label{fig:O3_O2_Op_flow}
\end{figure}
As we decrease $r_z$ away from 1, we observe that the primary fields of the O(3) fixed point split into a series of irreducible representations of O(2), following the branching rules of $\mathfrak{so}(3)\supset\mathfrak{so}(2)$(see Tab. \ref{tbl:O3_O2_branch}).

\begin{table}[!htb]
    \centering
    \setlength{\tabcolsep}{7pt}
    \caption{ The branch rule of $\mathfrak{so}(3)\supset\mathfrak{so}(2)$. 
    }
    \begin{tabular}{ccc|cc}
        \hline\hline
        $j$ &$\text{Dim}_{\mathrm{SO(3)}}$&$C_2^{\mathrm{SO(3)}}$&SO(2) Irep($j_z$) &$C_2^{\mathrm{SO(2)}}(j_z^2)$\\
        \hline
        0& 1 &0& $0$ &0 \\
        1& 3 &2& $0,\pm 1$ &0,1 \\
        2& 5 &6& $0,\pm 1,\pm 2$ &0,1,4 \\
        3& 7 &12& $0,\pm1,\pm2,\pm3$ &0,1,4,9 \\
        4& 9 &20& $0,\pm1,\pm2,\pm3,\pm4$ &0,1,4,9,16 \\
        \hline\hline
    \end{tabular}
    \label{tbl:O3_O2_branch}
\end{table}

From Fig. \ref{fig:extrapolation_gap_at_deltaz_-0.5}, we observe the absolute gaps of splitted states.
Some splitted states, such as $\phi_{[0]},T_{[1]},J_{[1]}^\mu,\cdots$, are scaled to be gapped and thus do not belong to the conformal operator spectrum of the O(2) fixed point (the superscript $[j_z]$ denotes the irreducible representations of O(2) group.), while others remain nearly gapless and become new primary operators at the O(2) fixed point($\phi_{[1]},T_{[2]},J_{[0]}^\mu,\cdots$).  
\begin{figure}
    \centering
    \includegraphics[width=0.32\linewidth]{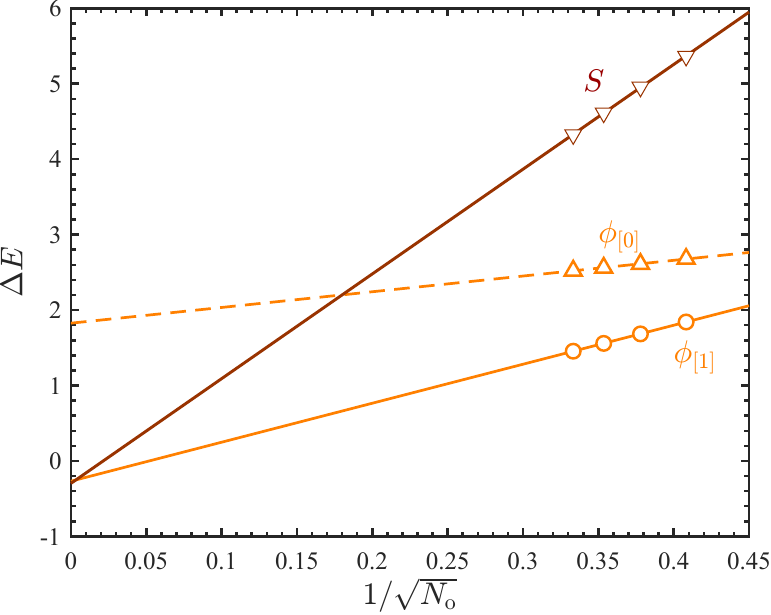}
    \includegraphics[width=0.32\linewidth]{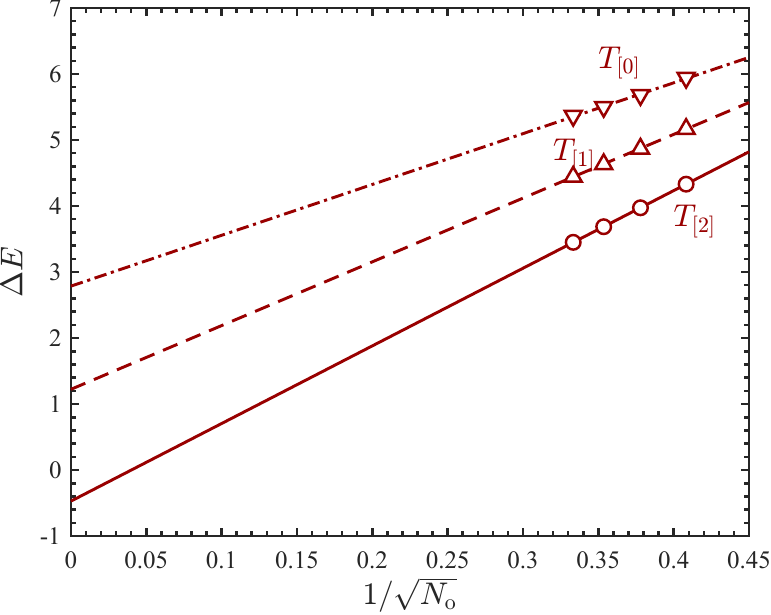}
    \includegraphics[width=0.32\linewidth]{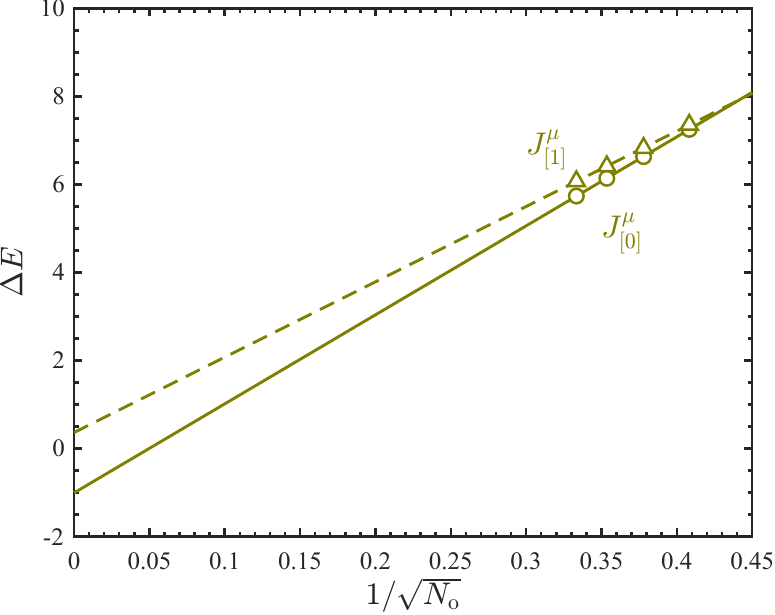}
    \caption{Finite size extrapolation of several energy gaps $\Delta E=E_n-E_0$ at the phase transition point when setting $r_z=-0.5$.}
    \label{fig:extrapolation_gap_at_deltaz_-0.5}
\end{figure}

In particular, we also observe an avoided level-crossing behavior between two singlet operators $S$ and $T_{[0]}$ with the same quantum numbers(Fig. \ref{fig:O3_O2_level_cross}(c)). In this case, we construct the following two operators:
\begin{equation}
    \Bar{S}(\boldsymbol{r})=\int d\boldsymbol{r^\prime}\delta(\boldsymbol{r}-\boldsymbol{r^\prime}) \boldsymbol{n}_1(\boldsymbol{r})\cdot\boldsymbol{n}_2(\boldsymbol{r^\prime}),
    \label{Seq:probe_Sbar}
\end{equation}
and
\begin{equation}
    T_{zz}(\boldsymbol{r})=\int d\boldsymbol{r^\prime}\delta(\boldsymbol{r}-\boldsymbol{r^\prime}) [n_1^z(\boldsymbol{r}) n_2^z(\boldsymbol{r^\prime}) - \frac{1}{3} \boldsymbol{n}_1(\boldsymbol{r}) \cdot \boldsymbol{n}_2(\boldsymbol{r^\prime})].
    \label{Seq:probe_Tzz}
\end{equation}
The $\Bar{S}$ operator is expected to have a finite overlap with the singlet $S$, but negligible overlap with $T_{[0]}$(split from O(3) rank-2 tensor $T$), while the reverse is true for $T_{zz}$.
By calculating the overlap between the two candidate excited states and the two probing operators(Eq. \ref{Seq:probe_Sbar} and \ref{Seq:probe_Tzz}) acting on the ground state, we find that the relative ordering of the two excited states reverses roughly when $r_z < 0$(see Fig. \ref{fig:O3_O2_level_cross}(a), (b) and (c)). This indicates that an avoided level crossing also occurs along the flow from the O(3) Heisenberg fixed point to the O(2) XY fixed point.

\begin{figure}
    \centering
    \includegraphics[width=0.32\linewidth]{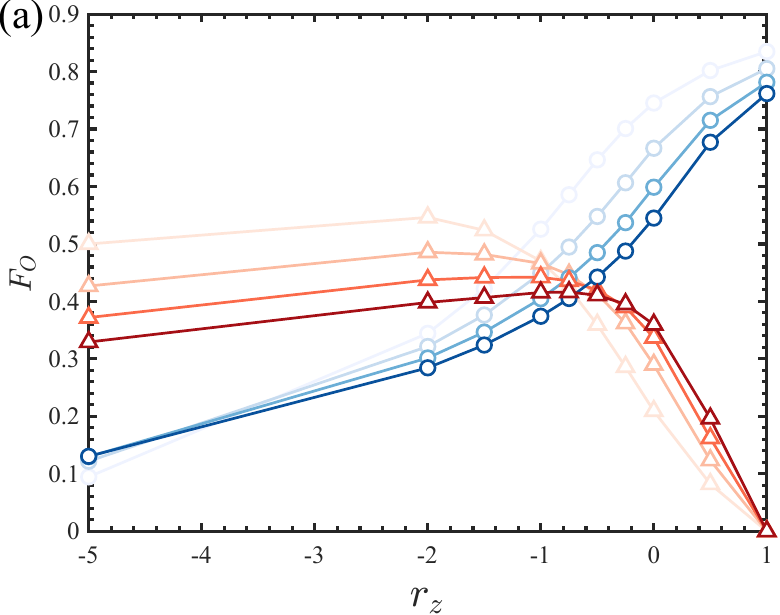}
    \includegraphics[width=0.32\linewidth]{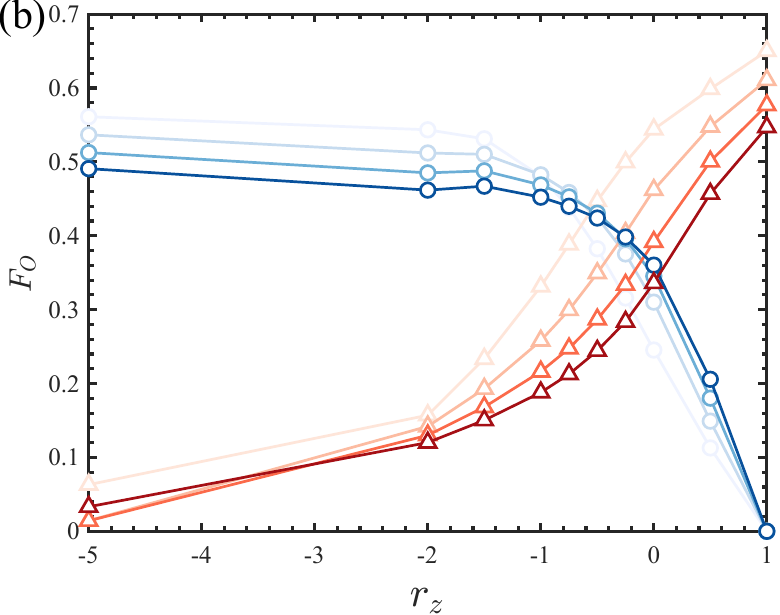}
    \includegraphics[width=0.32\linewidth]{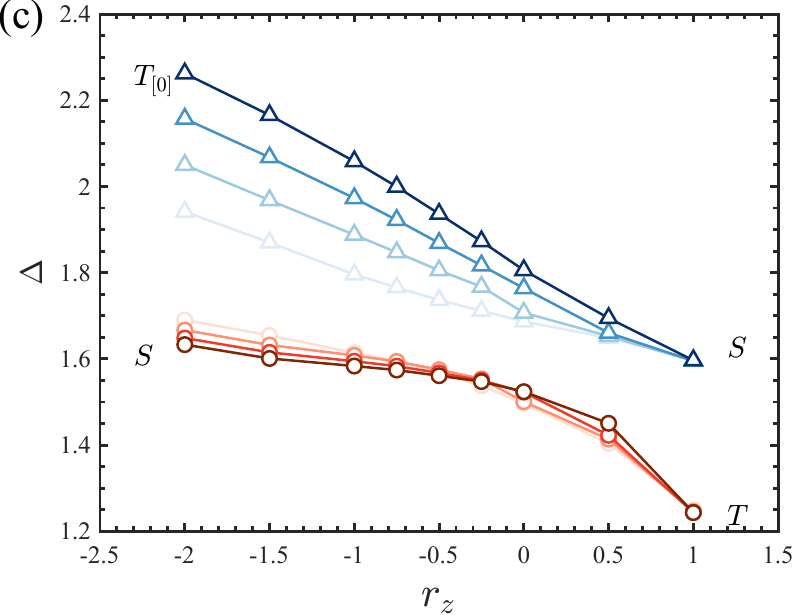}
    \caption{Level crossing behavior of $T_{[0]}$ and $S$ during the flow from the O(3) to the O(2) fixed point. Here $F_O=\langle (S, T_{[0]}) |O | I\rangle/|O|I\rangle|$ (a) $O = T_{zz}$; (b) $O = \bar{S}$.  The darkest color corresponds to results from the largest system size $N_\text{o}=9$. (c) Evolution of scaling dimensions of the two lowest singlets relating to $S$
and $T_{[0]}$, respectively. }
    \label{fig:O3_O2_level_cross}
\end{figure}

 Notably, in comparison to the SO(5) to O(4) case discussed in the main text, the \( \Bar{S} \sim \boldsymbol{\vec{n}}^2 \) operator constructed analogously for the putative SO(5) and O(4) fixed points turns out not to be a good approximation. 
 Beacuse the overlaps between $T_{[0,0]}$ or $S$ remain small throughout the SO(5)$\rightarrow$ O(4) flow. This suggests that the precise operator content of the relevant singlet $S$ at the approximate SO(5) and O(4) fixed points remains unclear, potentially due to the corresponding complex fixed points lying a finite distance away from the real axis in the complex plane. Understanding the nature of this relevant $S$ operator may help in identifying a genuine DQCP fixed point.

\begin{figure}[t]
\includegraphics[width=0.48\linewidth]{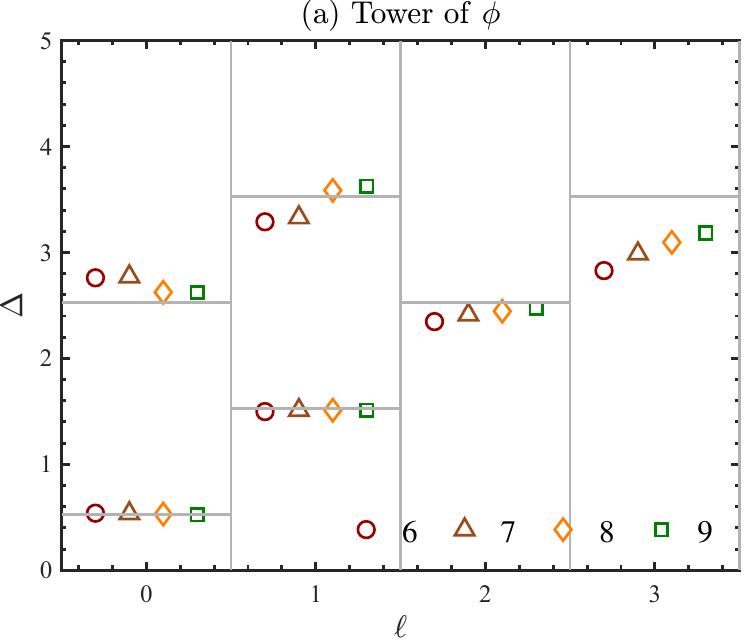}
\includegraphics[width=0.48\linewidth]{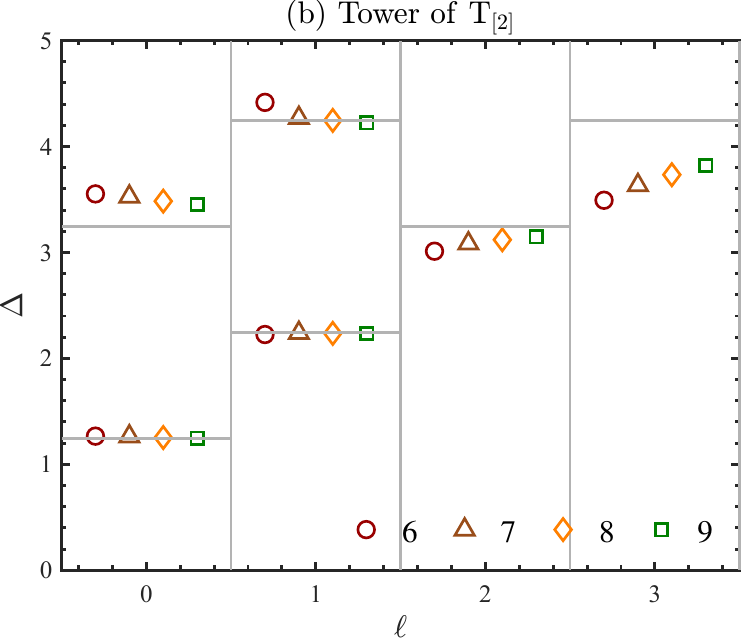}
\includegraphics[width=0.48\linewidth]{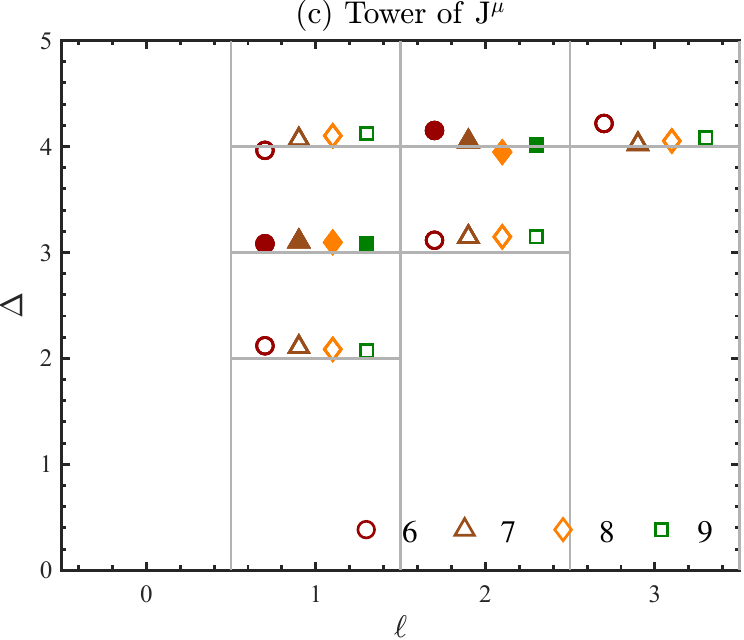}
\includegraphics[width=0.48\linewidth]{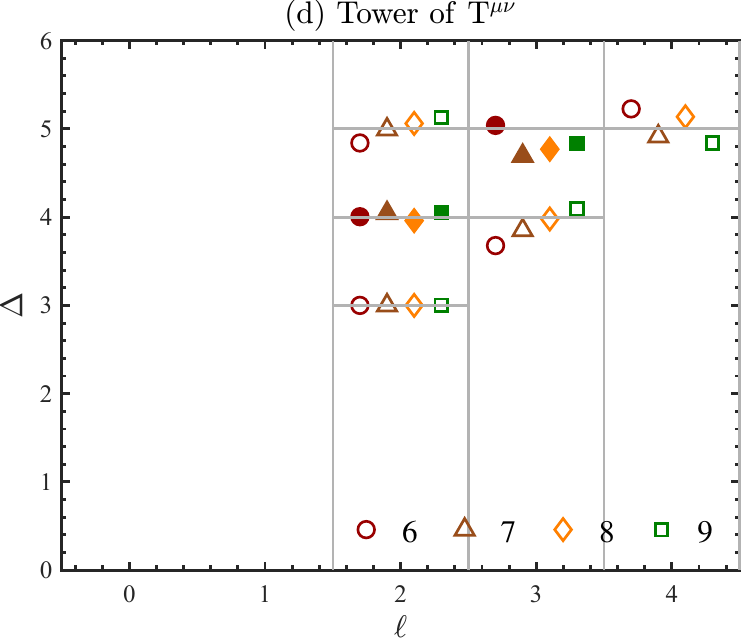}
	\caption{ The operator spectra of the conformal multiplets for  (a) $\phi_{[1]}$, (b) $T_{[2]}$, (c) $J^\mu$ and (d) $T^{\mu\nu}$. The horizontal grey lines denote the anticipated values from the conformal symmetry. These data are calculated at $r_z=-0.5$. Various symbols stand for the system size with orbitals 6-9. 
    } 
	\label{fig:O2Tower}
\end{figure}

Finally, we show the operator spectrum extracted at the critical point for $r_z=-0.5$ in Fig. \ref{fig:O2Tower} and several scaling dimensions of primary operators calculated under finite-size systems are listed in Tab. \ref{tbl:O2_Delta}. These results are in close agreement with the conformal data of the O(2) Wilson-Fisher fixed point obtained via the bootstrap approach. However, further efforts are needed to locate the O(2) fixed point by tuning parameters to suppress the influence of irrelevant operators and thereby reduce finite-size effects.

In conclusion, the overall features of RG flow from O(3) $\rightarrow$ O(2) Wilson-Fisher fixed point are very close to the case SO(5)$\rightarrow$O(4) case in the main text. We believe the findings here, such as operator decomposition and avoided level crossings, are general to the RG flows. The fuzzy sphere method used in this work could advance future studies on similar topics.   

\begin{table}[t!]
    \centering
    \caption{ Operator content of Wilson-Fisher O(2) fixed point.
    The scaling dimension and quantum numbers for the lowest lying primary operators obtained from state-operator correspondence at different system sizes $N_\mathrm{o}$ with $r_z=-0.5$. The transition point $h_c$ is determined by the curve crossing point analysis of Binder ratio.  }\label{STab:O2_deltaz_m0.5}
    \setlength{\tabcolsep}{4.5pt}
    \begin{tabular}{c|c|c|ccc|cccc}
        \hline\hline
        \multirow{2}{*}{$\epsilon$-exp \cite{HENRIKSSON2023}} &\multirow{2}{*}{Bootstrap\cite{bootstrap_rmp}} & \multirow{2}{*}{Op.} & &$N_\mathrm{o}$ &  &9&8&7&6\\
        \cline{4-10}
        &  & &$\ell$ & $\mathcal{P}$ & Rep.  & &$\Delta$ &&  \\\hline
        $0.5$& $0.519088$ & $\phi$                  & $0$ & $-$ & ${[1] }$ &  $0.5251$ & $0.5296$ & $0.5353$ & $0.5384$  \\
       $1.2$ & $1.23629$ & $T_{[2]}$                     & $0$ & $+$ & ${[2]}$ & $1.2450$ & $1.2526$ & $1.2627$ & $1.2658$  \\
       $2$ & $2$ & $J_{[0]}^\mu$                 & $1$ & $-$ & ${[0]}$ &  $2.0694$ & $2.0854$ & $2.1069$ & $2.1186$  \\
       $1.4$ & $1.51136$ & $S$                & $0$ & $+$ & ${[0]}$ &  $1.5609$ & $1.5675$ & $1.5472$ & $1.5103$  \\
        $2.8$ & -& $H^m_{[2]}$         & $1$ & $+$ & ${[2] }$ &  $3.0410$ & $3.0364$ & $3.0384$ & $3.0242$ \\
        $3$ & $3$ & ${T}^{\mu\nu}$  & $2$ & $+$ & ${[0] }$ &  $3.000$ & $3.000$ & $3.000$ & $3.000$  \\ 
       $2.1$ & $2.1086$& $t_{[3]}^3$     & $0$ & $-$ & ${[3]}$&  $2.1216$ & $2.1320$ & $2.1461$ & $2.1476$  \\
       $3.2$ &$3.111535$ & $t_{[4]}^4$     & $0$ & $+$ & ${[4]}$&  $3.1347$ & $3.1478$ & $3.1662$ & $3.1650$  \\
       $5.7$ &$-$ & ${\phi^\prime}$            & $0$ & $-$ & ${[1] }$ &  $4.2746$ & $4.2629$ & $4.2493$ & $4.1913$ \\
        \hline\hline
    \end{tabular}
    \label{tbl:O2_Delta}
\end{table}

\end{widetext}

%
\end{document}